\documentclass[twocolumn,pra,amssymb,amsmath,superscriptaddress]{revtex4-2}
\usepackage{braket}
\usepackage{tikz}
\usepackage{multirow}
\usepackage{booktabs}
\usepackage[export]{adjustbox}
\usepackage{graphicx}
\usepackage{afterpage}
\usepackage{amsthm}

\usepackage{xcolor}
\usepackage{hyperref}
\hypersetup{
	colorlinks=true,
	linkcolor=blue,
	filecolor=magenta,      
	urlcolor=blue,
	citecolor=red
}
\usepackage[english]{babel}
\usepackage[normalem]{ulem}
\usepackage{soul}
\usepackage{dcolumn}
\usepackage{bm}
\usepackage{mathtools}
\DeclarePairedDelimiter{\ceil}{\lceil}{\rceil}
\newtheorem{theorem}{Theorem}
\begin{document}
	
	\preprint{APS/123-QED}
	
	\title{Subexponential rate versus distance with time-multiplexed quantum repeaters}
	
	\author{Prajit Dhara}
	\author{Ashlesha Patil}
	\affiliation{Wyant College of Optical Sciences, The University of Arizona, Tucson, Arizona 85721 USA}
	\author{Hari Krovi}
	\affiliation
	{Raytheon BBN Technologies, 10 Moulton Street, Cambridge, Massachusetts 02138 USA}
	\author{Saikat Guha}
		\email{saikat@optics.arizona.edu}
	\affiliation{Wyant College of Optical Sciences, The University of Arizona, Tucson, Arizona 85721 USA}
	
	\begin{abstract}
	
Quantum communications capacity using direct transmission over length-$L$ optical fiber scales as $R \sim e^{-\alpha L}$, where $\alpha$ is the fiber's loss coefficient.  The rate achieved using a linear chain of quantum repeaters equipped with quantum memories, probabilistic Bell state measurements (BSMs) and switches used for spatial multiplexing, but no quantum error correction, was shown to surpass the direct-transmission capacity. However, this rate still decays exponentially with the end-to-end distance, viz., $R \sim e^{-s{\alpha L}}$, with $s < 1$. We show that the introduction of temporal multiplexing---i.e., the ability to perform BSMs among qubits at a repeater node that were successfully entangled with qubits at distinct neighboring nodes at {\em different} time steps---leads to a subexponential rate-vs.-distance scaling, i.e., $R \sim e^{-t\sqrt{\alpha L}}$, which is not attainable with just spatial or spectral multiplexing. We evaluate analytical upper and lower bounds to this rate, and obtain the exact rate by numerically optimizing the time-multiplexing block length and the number of repeater nodes. We further demonstrate that incorporating losses in the optical switches used to implement time multiplexing degrades the rate-vs.-distance performance, eventually falling back to exponential scaling for very lossy switches. We also examine models for quantum memory decoherence and describe optimal regimes of operation to preserve the desired boost from temporal multiplexing. Quantum memory decoherence is seen to be more detrimental to the repeater's performance over switching losses. 
	\end{abstract}

	\maketitle
	
	
	\section{Introduction}
	Shared entanglement is the basic building block for a wide variety of quantum information based protocols such as distributed quantum computing~\cite{Van_Meter2016-nn}, entanglement enhanced sensing~\cite{Proctor2018-bp,Zhuang2018-zu}, quantum key distribution~\cite{Bennett1992-bm} and more. The maximum rate at which entanglement can be created across a pure-loss optical channel of transmissivity $\eta$ by direct photon transmission, and authenticated two-way public classical communication, is $\propto \eta$ entangled bits (ebit, or ideal two-qubit Bell states) per transmitted optical mode, when $\eta \ll 1$~\cite{Takeoka2014-bh}. The exact expression of the entanglement generation capacity, $R_{\text{direct}}(\eta)= -\log_2 (1-\eta) $ ebits per mode~\cite{pirandola2017a}, called the PLOB bound hereinafter, is $ \approx 1.44\eta $ for $ \eta\ll 1$. Loss in optical fiber scales exponentially with range $L$, i.e., $\eta = e^{-\alpha L}$. Therefore, over long distances of optical fiber, entanglement generation rate (and hence the quantum communication rate, e.g., using teleportation) must decay exponentially with $L$. The ebits/second rate is the aforesaid ebits/mode rate times a modes/second multiplier $W$ (hertz), which would be governed by the bandwidth (of the source, detector, and memory). However, unlike classical communications, the quantum communications rate cannot be increased by turning up the transmitted power. {\em Quantum repeaters} are a special-purpose quantum processor~\cite{briegel1998a}, which when deployed along the length of the end-to-end optical communication channel (free-space or fiber), can enable the generation of entanglement across the end users Alice and Bob at a rate higher than $R_{\text{direct}}(\eta)$, where $\eta$ is the effective transmissivity corresponding to the end-to-end Alice-to-Bob channel. 
	
	Repeater designs of various kinds---involving different choices of encoding the qubit in the photonic modes, different forms of memories, error correction codes, and operational modality (one-way vs.~two-way)---have been proposed, analyzed, and shown to outperform direct transmission~\cite{guha2015,pant2017,jiang2009,bratzik2014,sangouard2011a}. {\em One-way} quantum repeaters encode quantum data using forward {\em quantum} error correction, and the job of repeaters is to decode and re-encode~\cite{Rozpedek2020-pv,jiang2009}. {\em Two-way} repeaters generate and purify entanglement across short channel segments, and string them together using Bell state measurements (BSMs) at the repeater nodes, into end-to-end entanglement; potentially interspersed with additional rounds of purification across longer (multi-hop) channel segments~\cite{guha2015}. With a two-way repeater based on dual-rail photonic qubits, heralded quantum memories, and spatial multiplexing---i.e., switches that can let a repeater pick, in each time slot, a pair of Bell states successfully-established with each of its two neighbor across a set of $M > 1$ parallel channels, to connect with a single BSM---one can achieve an entanglement distribution rate that scales as $e^{-\alpha s L}$ ebits/mode, where $s\in (0,1)$~\cite{guha2015}  is a scaling factor set by the protocol and device parameters. Therefore, despite the rate scaling exponentially with end-to-end distance $L$, it can outperform the highest rate achieved with direct transmission over $M$ parallel channels, $1.44 M e^{-\alpha L}$ ebits/mode, at a large distance $L$. Similar performance can be achieved with quantum memories replaced with all-photonic cluster states that mimic the action of memories by providing quantum error correction against optical loss seen as a error mechanism on the dual rail photonic qubit~\cite{pant2017}.
	
In this paper, we consider a two-way quantum repeater architecture of the kind described above, but one that incorporates temporal multiplexing in addition to (or, in lieu of) spatial multiplexing. It must be noted that the effect of temporal multiplexing on the repeater's rate-vs.-distance performance is fundamentally different from spatial multiplexing, as it allows for the accrual of a greater number of qubits at a repeater node (entangled with qubits held at neighboring nodes) at the cost of a longer `wait time' before a BSM is performed at the repeater node, potentially connecting two Bell pairs generated across the neighboring links in two different time steps. These two effects counter each other to yield a subexponential rate, i.e., the end-to-end entanglement rate goes as $e^{-t \sqrt{\alpha L}}$ ebits/mode with $ t<1 $. This was first noted, as a scaling law, in~\cite{razavi2009}. Other proposals for time multiplexing propose the use of subsequent linear optical BSM attempts between the repeater nodes to enhance the rate of shared entanglement generation. A time-multiplexed scheme for shared entanglement generation between few nodes in various configurations of the associated hardware is analyzed in~\cite{Van_Dam2017-er}. A multistep entanglement length doubling scheme with multiplexed repeaters is proposed in~\cite{Collins2007-jl}. However a majority of the existing analyses are targeted for smaller repeater networks and do not address the subexponential rate scaling with distance. Specifically, existing literature does not consider time multiplexing as an optimizable parameter that potentially yields the improved scaling law reported in this paper. We develop this result rigorously, to yield an {\em exact} characterization of the repeater performance in the presence of losses (in coupling, fiber, detectors and switches) and memory decoherence, while optimizing the number of repeater nodes to deploy between Alice and Bob, and the optimal time-multiplexing block length, given the total end-to-end range $L$. We develop analytical lower bounds to this optimal rate that provide useful intuitions on how the achievable entanglement rate is affected by various device metrics. We find that when the switching losses or the memory decoherence increase beyond a threshold, the advantage afforded by time multiplexing is lost. The analysis in this paper is limited to the rate performance of the protocol and does not account for infidelities introduced in the ebit during any of the protocol steps (i.e.\ generation, transmission, storage or measurement). 
	
The paper is organized as follows. Section~\ref{section:architecture} describes the repeater architecture. The sub exponential bounds on the achievable entanglement generation rate are derived in Sec.~\ref{section:bounds}, highlighting the fundamental advantage of time multiplexing over spatial multiplexing alone. The practical limitations of a loss-prone switching network is demonstrated in Sec.~\ref{section:switching}. Further error models considering a `worst case' decoherence for quantum memories are studied in Sec.~\ref{section:decoherence}. More realistic decoherence models considering a variety of BSM scheduling protocols are examined in Sec.~\ref{section:avg_decoh}. Section~\ref{sec:conclusions} concludes the paper with thoughts on further extensions and applications of our results, e.g., to quantum networks that go beyond a line topology and that support multiple user pairs and groups.

\section{Quantum Repeater Architecture}
\label{section:architecture}

	\begin{figure*}[]
		\centering{\includegraphics[width=\linewidth]{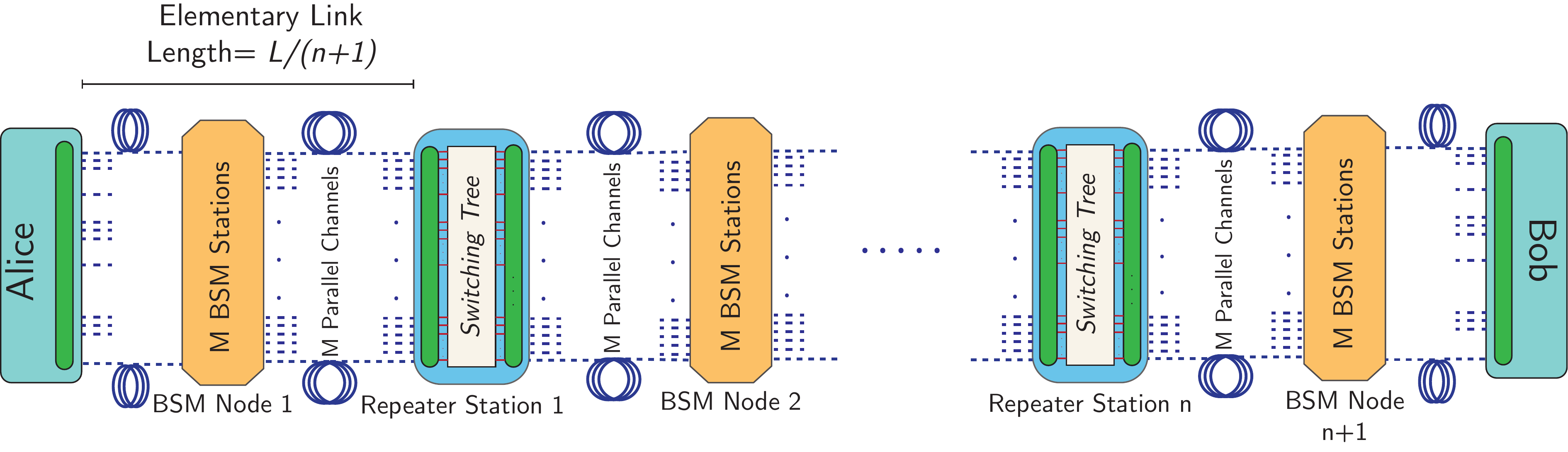}  }
		\caption{A linear repeater chain connecting two parties Alice and Bob. The end-to-end link is subdivided into $n+1$ (equal-length) elementary links. The light-blue boxes, on either end of an elementary link are quantum repeater (QR) stations, which are described in more detail in Fig.\ref{fig:repeater_station} and discussed in Section \ref{section:architecture}. The yellow boxes at the center of an elementary link contain $M$ linear-optical Bell state analyzers, each built with two beamsplitters and four single photon detectors. These perform BSM or entanglement swaps. Each elementary link has $M$ parallel channels, which could be optical fiber bundles, multi-mode optical fiber, multi-spatial-mode diffraction-limited near-field free-space channels, or wavelength-division multiplexed channels at $M$ frequencies on a single fiber. These $M$ parallel channels allow multiplexing to boost entanglement generation rates.}
		\label{fig:repeater_chain}
	\end{figure*}
\begin{figure}[h!]
		\centering{\includegraphics[scale=0.4]{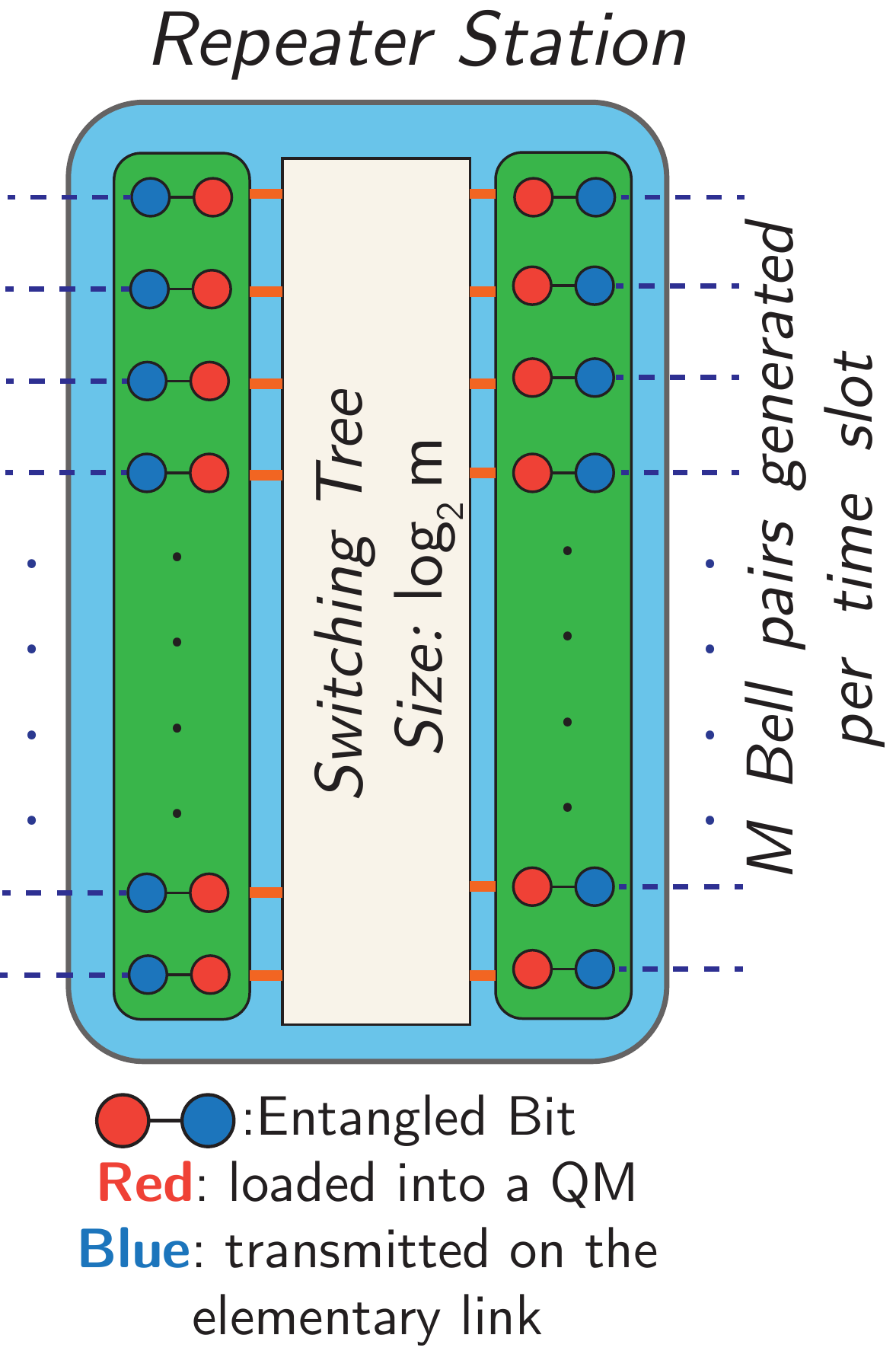}  }	
		\caption{Quantum repeater (QR) station consisting of $2M$ entangled qubit pair sources and a heralded quantum memory (QM) register with a large storage buffer. The sources generate two-qubit Bell states every $ \tau $ seconds. One of those qubits (denoted as a blue qubit) is a photonic qubit that is transmitted on one of the $2M$ channels: $M$ going to the right over one elementary link, and $M$ to the left over another elementary link. The red qubits are stored in the QM register. One could have photonic entanglement sources with one photonic qubit {\em loaded} onto the QM and the other transmitted. Alternatively, one could have memories that transmit a photon entangled with a qubit native to the memory's internal degree of freedom (e.g., a spin). At the end of a block of $m$ time slots, the QR station needs to perform a single BSM on two qubits of a total of $Mm$ qubits held in the QM register. Since switching across a total of up to $m$ time slots is involved, a switch array consisting of $ \log_2 m $ switches is needed. If the effective transmission corresponding to the per-switch loss is $\lambda_t$, the BSM success probability $q$ thus would be multiplied by $\lambda_t^{\log_2m}$. If the memories can perform all-to-all quantum logic asynchronously, for example as in Ref.~\cite{bhaskar2020}, then the aforesaid multiplicative factor of $\lambda_t^{\log_2m}$ can be avoided.}
		\label{fig:repeater_station}
	\end{figure}

 
Let us consider a linear chain of quantum repeaters (QRs) between the two communicating parties Alice and Bob, where the distance $L$ between the two parties is divided into $n+1$ elementary links (shown in Fig.~\ref{fig:repeater_chain}). Every elementary link has a QR station (shown in Fig.~\ref{fig:repeater_station}) at both ends and BSM stations (shown by the yellow boxes in Fig.~\ref{fig:repeater_chain}) with linear optical circuits to perform Bell state measurements (BSMs), for entanglement swap, at its center. Every QR contains $2M$ entangled photon-pair sources, with each source `associated with' (i.e., dedicated to) one of the $M$ parallel (e.g., spatially, or spectrally-multiplexed) channels going out to its right-hand QR neighbor, or one of the $M$ channels going to its left-hand QR neighbor. Each source generates dual-rail-encoded Bell pairs at a fixed repetition rate (say, one pair of entangled photons every $\tau$ seconds). We assume ideal and deterministic entanglement sources in our analysis~\cite{dousse2010,huang2012}, hence the entanglement generation rate of each source is $1/\tau$ ebits/second~\footnote{One way to achieve near-deterministic entangled-photon pair sources is via multiplexing many sources. An alternative is to use quantum dot or other quantum emitter like sources.}. Let us begin with an assumption that each QR has access to a random-access quantum memory (QM) register consisting of a large number of {\em heralded} single-qubit memories that have arbitrarily long coherence times. We will quantify the actual requirements as to the number of memories in the register and required coherence times in Sec.~\ref{subsec:timing}. A heralded QM implies it raises a {\em success flag} when it successfully loads a (photonic) qubit into its native (e.g., atomic spin based) qubit storage form.  We assume that the loading process into the QM takes a negligible amount of time and there is perfect clock synchronization across all QR nodes. At the clock edge of each time slot, each of the $2M$ sources generates an entangled photon pair---a dual-rail (e.g., polarization-encoded) single-photon Bell state of the form $\ket{\Psi}=[|\bar{1},\bar{0}\rangle + |\bar{0},\bar{1}\rangle]/\sqrt{2}$---and loads one photon (the red-colored qubit illustrated in Fig.~\ref{fig:repeater_station}) into the QM register at that QR node while transmitting the other photon (the blue-colored qubit) into one of the outgoing $2M$ channels towards the center of one of the two elementary links. At the center of each elementary link sits an array of $M$ linear-optical Bell state analyzers, one for each of the $M$ parallel channels. Each performs a BSM between a pair of qubits flying in from opposite directions from two QR nodes. The protocol analysis in this section assumes that switching photons between the constituent QMs can be achieved with no loss. Additionally, we assume that the qubit stored in the QM does not undergo any decoherence or loss. These assumptions are relaxed in the analyses of Secs.~\ref{section:switching}-\ref{section:avg_decoh}.

Each photonic qubit undergoes lossy transmission to the center of the elementary link, thereby going through an effective transmittance $\lambda = e^{-\alpha L/[2(n+1)]}$, where $ \alpha $ (typically measured in dB/km) is the fiber's loss coefficient. We assume the most basic linear optical BSM circuit, which heralds successful entanglement swap with a probability of 50\%~\cite{lutkenhaus1999} when built with ideal devices (ideal 50-50 beamsplitters, and lossless and noiseless ideal single photon detectors). Therefore, in one given time slot, and across one of the $M$ parallel channels, an elementary link (henceforth referred to as a {\em link}) successfully establishes an entangled Bell state across a pair of qubits held at QM registers at two neighboring QR nodes, with probability $\mu \lambda^2 =\mu \exp[-\alpha L/(n+1)]$. Here $\mu = \eta_d^2/2$ combines the effects of the intrinsic $1/2$ success probability of a linear-optical BSM and the detector efficiencies of two (of the four) single photon detectors within the linear-optical BSM circuit whose `clicking' heralds the success of the BSM. At the end of each time slot, the one-bit success-failure information about each of the $M$ BSMs is sent back to both neighboring QR nodes via a separate error-free classical communications channel. Table~\ref{tab:protocol_parameters} summarizes the parameters associated with the proposed protocol.

\subsection{Time multiplexed BSMs}

Thus far, the repeater protocol sounds identical to the one analyzed in~\cite{guha2015}. Our protocol has an additional parameter: the time-multiplexing block length $m \ge 1$, which is known to all QR nodes. The protocol in Ref.~\cite{guha2015} is a special case of our protocol corresponding to $m=1$. The complete sequence of events described above---generation of entanglement, loading one qubit of each Bell state into the QR register, transmission to the middle of the link, linear-optic BSMs, and classical communication of  the swap outcomes---is repeated over $m$ time slots (i.e., $m\tau$ seconds), over all $M$ parallel channels in each elementary link. When a QR node has received success-failure information from each of the $Mm$ BSM attempts in the link on its right and the $Mm$ BSM attempts in the link on its left, it picks the first successful BSM on its right and the first successful BSM on its left, assuming there is at least one success on each side in that time block, and performs a BSM among the two (red) qubits held in the QM register corresponding to those successful links on either side of the QR. The QR nodes must pre-agree on a common ordering indexing the $Mm$ channels and time slots, so that {\em first success} in the above sentence is consistent. The QR nodes do not need to coordinate during the protocol run for this step, as their action (choice of which two red qubits in their QM register to perform a BSM on) is predicated only by the success-failure outcomes coming from the QR node's neighboring links. Each QR node thus performs a BSM on a pair of qubits held in its QM register, every $m\tau$ seconds. The rule for the BSM at the QR node will be modified slightly in Section \ref{section:decoherence} when non-idealities in the QM storage characteristics are accounted for. 

It is simple to see that the probability that at least one BSM in the $ m\times M $ attempts succeeds, is 
	\begin{equation}\label{eq:P_single}
		P = 1 -(1 -\mu \exp(-\alpha L/(n+1)) )^{M m} \,.
	\end{equation}
There is no fundamental limit to the success probability of the BSM for the qubits loaded in QMs. However, in order to account for any loss in loading the photonic qubit into the QM register, we allow for a sub-unity probability of successful swap, which we take as $ q $. Note that all of the above (i.e.\ the steps of the protocol until the QM BSM) could have been implemented by a QM register which {\em emits} photons every $\tau$ seconds that is entangled with an internal qubit, such as a spin (of a color center qubit register) or an ion in a trapped-ion qubit register. If so, we would not have to {\em load} the red qubit into a QM register. The red (matter) qubit would generate a blue (photonic) qubit entangled with it in a Bell state, in every time slot.

At the end of each $m\tau$ second block, when every QR node (simultaneously) performs a BSM in the QM register, if (1) each of the $n+1$ links had heralded at least one Bell state across it (whose probability is $P^{n+1}$), and (2) all the $n$ BSMs at the QR nodes succeeded (whose probability is $q^n$), the end users Alice and Bob would be delivered one ebit worth of shared entanglement. The end-to-end entanglement generation rate, in terms of our two design parameters: $n$ (number of QR nodes) and $m$ (time-multiplexing block length), is therefore given by: 
	\begin{eqnarray}
		R_{m,n}(L)&=& \frac{{P}^{n+1}q^n}{m\tau} \,{\text{ebits/second}}\nonumber \\
		&=&\frac{\left(1-\left(1-\mu e^{-\alpha L/(n+1)}\right)^{Mm}\right)^{n+1} \,q^n}{m \tau}.
		\label{eq:rateeqn}
	\end{eqnarray}
The analysis above assumes photon loss as the only detriment. In other words, a `success' of all the probabilistic steps in the protocol ensures an ideal Bell pair (one ebit of entanglement) to be delivered to Alice and Bob. This is why entanglement rate is simply the overall success probability divided by the effective repetition period. If the end-to-end two-qubit entangled state produced (upon overall success) is not a perfect two-qubit Bell state, the asymptotic achievable rate is given by the probability of success times the distillable entanglement per copy of that imperfect state. A lower bound to that distillable entanglement is given by the reverse coherent information (RCI). We may then employ entanglement distillation as an additional layer to generate high-fidelity end-to-end Bell pairs. Using block-distillation codes, the repeater nodes at the end points of the elementary link could convert $N$ sub-unity Fidelity entangled pairs into $K$ unit-Fidelity Bell states, only using local operations and classical communications (LOCC). This will require $N$ and $K$ to grow large, which will increase the latency of the protocol. However, the rate $K/N$ of that entanglement distillation code will be a constant determined by the distillable entanglement per copy of the imperfect Bell state~\cite{Bennett1996-fx}.

\begin{table}
	\caption{Parameters associated with the time multiplexed repeater network and protocol discussed in Section~\ref{section:architecture}. Recall that in terms of the detector efficiency($ \eta_d $), $ \mu=\eta_d^2/2 $.}
	\begin{tabular}{lc}
		\toprule
		\textrm{Parameters} & \textrm{Symbol} \\
		\midrule
		Fiber loss coefficient  & $\alpha$ \\
		Total network length  & $L$ \\
		Source repetition time & $ \tau $\\
		No. of quantum repeaters & $ n $ \\
		No. of parallel spatial/spectral channels & $ M $ \\
		Time multiplexing block length & $ m $ \\
		Quantum memory entanglement swap prob. & $ q $\\
		Linear-optical BSM efficiency & $\mu$\\
		\bottomrule
	\end{tabular}
	\label{tab:protocol_parameters}
\end{table}

\subsection{Timing diagram and QM requirements}
\label{subsec:timing}
We will now derive the minimum number of qubits that the QM register at a QR node needs to be able to store at any given time, and for how long (coherence time). The reader may have noticed that there is a finite latency to our protocol meaning that there is a delay between when the protocol is initiated and when the information about the BSM success-failure outcomes from the first $m$ time slots has arrived back at the QR node, and hence the QRs are ready to perform their BSMs and thus the first ebits begin to be delivered to Alice and Bob. This initial latency $T_l = T_1 + T_2$ has two parts. The first is $T_1$: the time light takes to travel from a QR node to the middle of the elementary link and back. $T_1 = L/(n+1)c_{\rm fib}=j\tau$, where $c_{\text{fib}}$ is the speed of light in optical fiber. Let us express it as $T_1 = j\tau$ (expressed as a multiple of the repetition period). The second is $T_2 = m\tau$, the wait time until the end of the $(m \tau)$th time slot, when the QR node has generated and transmitted the $ 2mM $ qubits for the intermediate linear-optical BSMs. Therefore, the initial latency (generation + qubit transmission + classical communication to the QR node) can be expressed as:
\begin{equation}
T_{l}=m\tau+\frac{L}{(n+1)c_{\text{fib}}}.
\label{eq:t_latency}
\end{equation}
Note that we do not account for the interaction and detection time associated with the linear-optical BSM when stating our latency. With realistic devices, the latency is slightly modified by the nature of the BSM circuit chosen. Nevertheless, this amount of time would be quite small compared with either $ T_1 $ or $ T_2 $, and can effectively be ignored.

\begin{figure}[h]
		\centering{\includegraphics[width=\linewidth]{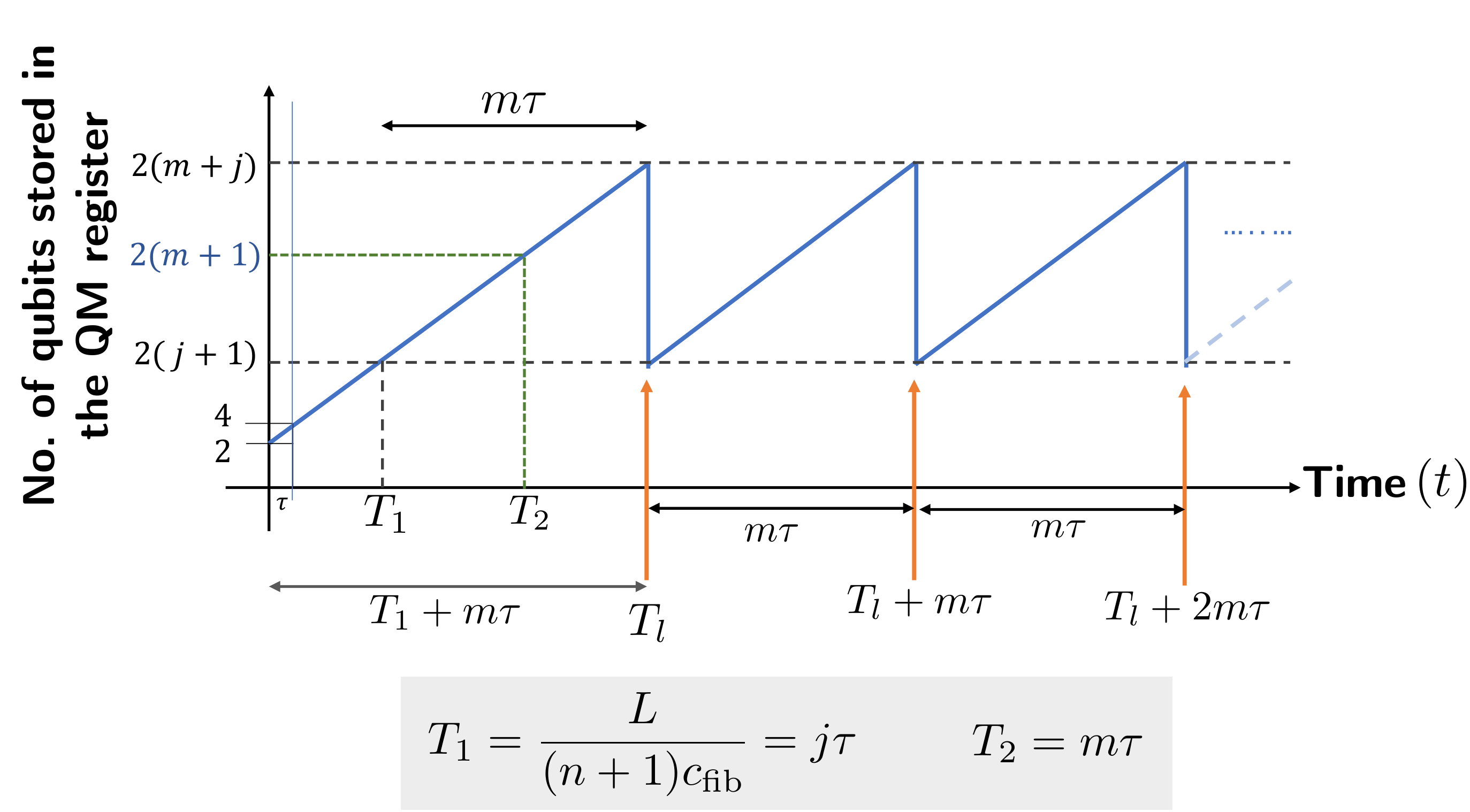}  }	
		\caption{Timing diagram showing the number of qubits stored in the QM register at a QR node as a function of time, the time-multiplexing block length $m$ and the number of repeaters $n$, with $M=1$ (scaling the $ y $-axis by $ M $ to account for spatial multiplexing). The initial latency time $T_l$ comprises two time segments. $T_1$ is  the time of flight for the transmitted photonic qubit to reach the middle of the elementary link, and the classical information about the BSM success to travel from the middle of the elementary link back to the QR node. $T_2 = m \tau$ is the additional time the QR node needs to wait until its first QM BSM is completed. After the initial latency of $ T_l=T_1+T_2 $, there are always $ 2mM $ memory slots that are occupied for shared entanglement to be generated between Alice and Bob in every $m\tau$ second interval.}
		\label{fig:timing_diagram}
	\end{figure}

A timing diagram for the protocol is shown in Fig.~\ref{fig:timing_diagram}. The vertical axis shows the number of qubits actually stored in the QM register at a QR node as a function of time. Let us assume $M=1$ for simplicity.  For $M>1$, the number of QMs occupied will be trivially multiplied by $M$ i.e.\ the axis can be scaled with $ M $. 
	
Two qubits are loaded in the QM register in each time slot. Thus, upto time $ t=T_1 $, we have loaded $ 2(T_1/\tau+1)=2(j+1) $ qubits. The orange arrows in Fig.~\ref{fig:timing_diagram} mark the times at which the QM BSMs are performed. The sharp drop at one of those time-slot boundaries corresponds to the QM register discarding $2m$ qubits, and loading $2$ fresh qubits. Hence, the QM register occupancy at these times (generally expressed as $ t_{meas}=T_l+(n-1)T_2$ where $ n\in \mathbb{Z}^+ $) is given by $ 2(m+j)-2m +2 =2(j+1) $.

 It is clear from this diagram that the latency in Eq.~\eqref{eq:t_latency} also determines (1) the buffer length (number of qubits) of the QM register required to support this protocol, and (2) the coherence time requirement, i.e., how long a qubit must be stored in the QM register from when it was generated (entangled with its photonic-qubit pair) until when it is measured (in a BSM, with another qubit in the QM register).

The minimum coherence time required is $T_{c,{\rm min}} = T_l-\tau$, because the last slot is filled as the measurement is made, i.e., the QM register being operated in a first-in first-out mode. Therefore,
	\begin{align}
		T_{c,{\rm min}}=\frac{L}{(n+1)c_{\text{fib}}} +(m-1)\tau .
		\label{eq:tmin}
	\end{align}
It is simple to see from the timing diagram that the minimum size of the QM register (buffer length) is:
	\begin{align}
		N_{\text{mem,min}}= 2\left(m+ \ceil[\bigg] {\frac{L}{(n+1)c_{\text{fib}}\tau}} \right) \,{\text{qubits}}.
		\label{eq:memmin}
	\end{align}
The optimal rate bounds that are presented in subsequent sections implicitly assume that the resource requirements in Eqs.~\eqref{eq:tmin} and~\eqref{eq:memmin} are met. The entanglement generation rate would be adversely affected if the available resources (length of QM register and/or the coherence time) are less than the above specified requirements. However the evaluation of such performance degradation will not be done in this paper. A detailed analysis in the context of time multiplexing using trapped ion quantum repeater modules has been presented in Ref.~\cite{Dhara2021-jo}.
	
\section{ Theoretical Analysis: Rate benefit of temporal multiplexing}
	\label{section:bounds}
	
\subsection{Main results}
The end-to-end entanglement rate is obtained by optimizing the rate equation in Eq.~\eqref{eq:rateeqn} with respect to the design parameters $ m $ and $ n $. The optimal rate-vs.-distance envelope, i.e., the rate $R(L) = \max_{m,n} R_{m,n}(L)$ is hard to derive. We derive upper and lower bounds to the actual envelope $R(L)$ that have similar mathematical characteristics to $R(L)$, to gain insight into the performance.

	\begin{theorem}
		The rate-vs.-distance envelope $ R(L) $  is bounded by two subexponential rate laws: $ R^{\text{UB}}(L) \ge R(L) \ge R^{\text{LB}}(L) $, which are given as follows:
		\begin{align}
			R^{\text{UB}} (L)&:=\frac{M \mu}{q \tau} e^{-\left(2 \sqrt{\log \left(1 / q\right)}\right) \sqrt{\alpha L}},\, {\text{and}} \label{eq:lowerbound}\\
			R^{\text{LB}} (L)&:=\frac{M \mu}{q \tau} e^{-\left(2 \sqrt{\log \left(1 / q\left(1-1/e\right)\right)}\right) \sqrt{\alpha L}}. \label{eq:upperbound}
		\end{align}
		\label{theorem:bounds}
	\end{theorem}
	\begin{proof}
		See Appendix \ref{appendix:bound} for a detailed proof. 
	\end{proof}
	
\noindent Theorem~\ref{theorem:bounds} holds even with $M=1$. In other words, in order to get the subexponential rate-vs.-distance scaling  when QRs are equipped with the machinery to perform time multiplexed BSMs, one does not need parallel channels, i.e., no spatial and/or spectral multiplexing is required. Just being able to mix and match successful entanglement attempts ( $ 2m $ attempts in this case) across neighboring links generated in different time slots is all one needs. In the expressions in Eqs.~\eqref{eq:lowerbound} and~\eqref{eq:upperbound}, the pre-factor $M/\tau$ is the number of transmitted modes per second, which when multiplied by the probability of successfully generating an ebit, gives the generation or distribution rate.
	
\begin{figure}[h!]
	\centering{\includegraphics[width=\linewidth]{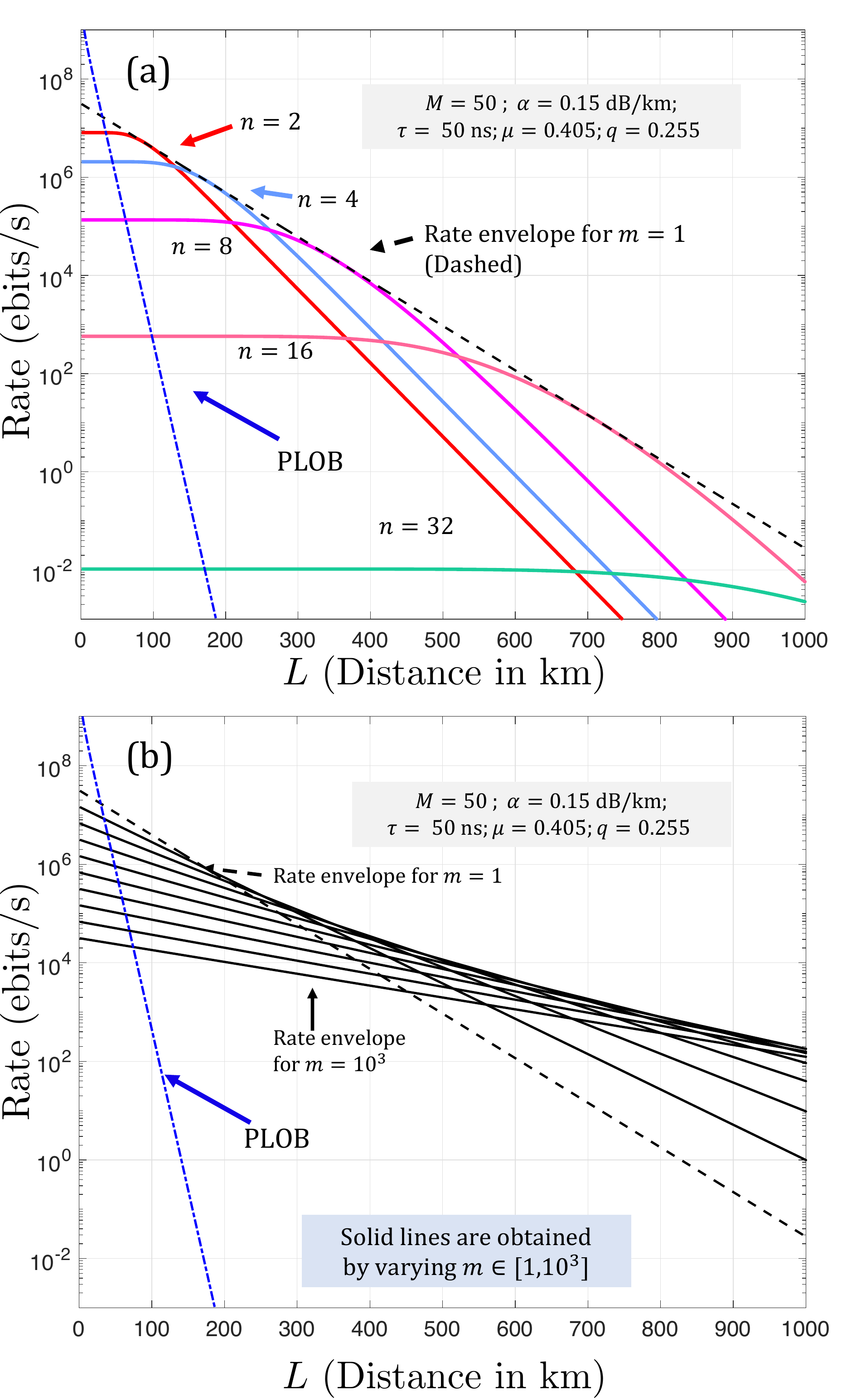}  }	
	\caption{ Demonstration of the emergence of the subexponential rate-vs.-distance envelope from time multiplexing. We assume repeaters with  $M=50; \alpha \equiv 0.15 \text{ dB/km}, \tau= 50 \text{ ns}, \mu=0.405, q=0.255$. (a) Exponential rate envelope (black dashed line) for a purely spatially multiplexed ($ m=1 $) repeater architecture with increasing number of repeaters in the chain (various colored lines; value of $ n $ marked). (b) Rate-distance envelopes for increasing values of $ m $ (black lines), with $n$ optimized at any given $L$. The outer envelope of the rate-distance envelopes as $m$ is varied, is seen to have a subexponential scaling, and is bounded by Eqs.~\eqref{eq:lowerbound} and~\eqref{eq:upperbound}.} 
	\label{fig:temporal_trend}
\end{figure}

If communicating directly over $M$ parallel channels, without the aid of any repeaters, the optimal rate comes from the PLOB bound of $-\log_2(1-\eta)$ ebits/mode:
\begin{eqnarray}
R_{\text{direct}}(L) &=& \frac{M}{\tau} \, \log_2\left(\frac{1}{1-\eta}\right)\\
&\approx& \frac{M}{\tau \ln 2} e^{-\alpha L},\, {\text{for}}\, \eta = e^{-\alpha L} \ll 1. 
\label{eq:Rdirect}
\end{eqnarray}
If time multiplexing wasn't used, and no parallel channels were employed either, i.e., $m = M = 1$, then it is not possible to outperform $R_{\text{direct}}(L)$. If no time multiplexing was used, and only spatial and/or spectral multiplexing were used ($m = 1, M > 1$),then  the end-to-end entanglement rate $R(L)$ is given by~\cite{guha2015}
\begin{equation}
R(L) = \frac{1}{q\tau} e^{-s \alpha L},
\label{eq:rate_spatialonly}
\end{equation}
where the exponent $s < 1$, and is given by
\begin{equation}
s = \log\left[q(1-(1-\mu z)^M)\right]/ \log z,
\end{equation}
with $z$ being the solution of the following transcendental equation $[1-(1-\mu z)^M] \log[q(1-(1-\mu z)^M)] = \mu M z(\log z) (1-\mu z)^{M-1}$~\cite{guha2015}. A more useful expression for the entanglement rate is an upper bound $R^{\rm UB}(L)$ to $R(L)$, from which the dependence of $M$ and $q$ can be seen more explicitly:
\begin{equation}
R^{\rm UB}(L) = \frac{1}{q\tau} e^{-u \alpha L},
\label{eq:rate_spatialonlyUB}
\end{equation}
where the exponent $u$ is given by~\cite{guha2015} as,
\begin{equation}
u = \frac{\log(1/q)}{\log(\mu M)}.
\end{equation}
For this upper bound to outperform the linear scaling $R \propto \eta$ of $R_{\text{direct}}(L)$, one needs $u < 1$, which implies $M > 1/(q\mu)$, and is intuitively consistent. One other important observation is that in Eqs.~\eqref{eq:rate_spatialonly} and~\eqref{eq:rate_spatialonlyUB}, the modes per second multiplier $(M/\tau)$ does {\em not} factor out of the rate expression, as it does for $R_{\text{direct}}(L)$ as seen in Eq.~\eqref{eq:Rdirect} and for the rate expression that includes optimized time multiplexing, as seen in Eqs.~\eqref{eq:lowerbound} and~\eqref{eq:upperbound}. Figure~\ref{fig:temporal_trend} demonstrates the emergence of the $ R(L) =O(e^{-\beta\sqrt{\alpha L}})$ scaling for temporal multiplexing. Starting with a $R(L) = O(e^{-\gamma{\alpha L}})$ envelope (black dashed line) in Fig.~\ref{fig:temporal_trend} (a) for $ m=1 $, varying the value of $ m $  in Fig.~\ref{fig:temporal_trend} (b) yields a set of exponentially-decaying rate-distance envelopes with different values of $ \gamma $ (black lines). The subexponential rate-vs.-distance scaling is traced by the upper edge of this set of envelopes.

The reason for why temporal multiplexing leads to $R(L) = O(e^{-\beta\sqrt{\alpha L}})$ as opposed to spatial multiplexing only achieving $R(L) = O(e^{-\gamma{\alpha L}})$, for constants $\beta$ and $\gamma$, can be seen from the proof of Theorem~\ref{theorem:bounds} in Appendix~\ref{appendix:bound}. Unlike spatial multiplexing, which boosts the probability of generating at least one successful entangled pair across an elementary link by $M > 1$ parallel entanglement attempts, temporal multiplexing can boost that probability much more, seemingly arbitrarily so by increasing $m$ as much as one would like. However, by increasing $m$, the effective time step increases from $\tau$ to $m \tau$, which degrades the rate, as $m\tau$ appears in the denominator of the rate equation~\eqref{eq:rateeqn}. However, the boost in the success probability of the link, along with the optimization of $n$, the number of QR nodes, outdoes the aforesaid degradation.

Intuitively, the subexponential scaling is closely tied to the fact that the protocol allows the time-multiplexing block length to be chosen optimally given the Alice-to-Bob range $ L $. If $ m $ is held fixed, then the rate scaling is still exponential, as this corresponds to using an effective but \emph{fixed} multiplexing size $ M'=M\times m $. If $ m $ is held fixed, but $ n $ can be optimized, then we revert back to the scheme of Ref.~\cite{guha2015}, which does not give an $ n $-dependent improvement in the exponent of the aforesaid exponential scaling. Now, if we allow $ m $  to be optimally chosen based on the repeater nodes’ knowledge of $ L $, then the rate scaling can be shown to scale subexponentially with $ L $.

\subsection{Discussion of device imperfections}
			
{\em Switching losses}---Although temporal multiplexing yields a better rate-loss scaling than spatial multiplexing, there are several requirements for the protocol to work. The most important is an optical switching network to connect successful elementary links, especially if the qubits in the QM register must be read out in the photonic domain before the BSM at the QR node. As the time multiplexing block length ($m$) is a parameter that we optimize over, the size of the optical switching network must also grow with it. A switching tree of size $\log_2 m$, which connects $m$ possible slots in which successes may occur, to a single output line, is the optimal choice in terms of using the least number of switches. This is shown as the white box in Fig.~\ref{fig:repeater_station}. If the loss per switch is quantified by an effective transmissivity $\lambda_t$, then the total loss of the switching array is $\lambda_t^{\log_2 m}$.
	
{\em QM coherence time}---Two-way quantum repeater protocols require QMs in some form. The current design assumes that QMs are available at every repeater station which do not maximally decohere before the information about the BSM is received at the repeater station. Therefore, it is important to note that the minimum coherence time is limited by the time it takes light to travel over the length of the elementary link, or $\tau_{\rm coh}> {L}/{\left[(n+1)c_{\text{fib.}}\right]}$ is the minimum requirement for the repeater chain to be able to generate shared entanglement. The exact coherence time requirements, as analyzed in Sec.~\ref{subsec:timing} is given by $ {L}/\left[{(n+1)c_{\text{fib}}}\right] +(m-1)\tau $.
	
{\em QM decoherence}---The degradation of the qubit held in the QM is another practical consideration that must be carefully analyzed. To simplify our analysis, we assume that the qubit is lost when it decoheres sitting in the QM register. This simply means that if the probability that a qubit does not decohere in time $\tau'$ is denoted as $p_0$, then after $k\tau'$ seconds, the probability is $p_0^k$. Furthermore, for a protocol that employs temporal multiplexing, one must take into account time-dependent decoherence: The two qubits on which the QR node performs the BSM, may not have suffered the same amount of decoherence as they may have been entangled across their respective elementary links at different time slots.

\section{Effect of switching losses}
	\label{section:switching}
	The parameter $\lambda_t$ quantifying per-switch loss, which is $10\log_{10}(1/\lambda_t)$ dB of loss, can be thought of as the probability with which a single switch succeeds in switching a photon (as opposed to losing it). Therefore, it can be subsumed in the BSM success probability $q$ at the QR node, which must be modified as: 
	\begin{align}
		q\rightarrow q \times \lambda^{\log_2 m}_t, 
		\label{eqn:switch_modif}
	\end{align}
	where $  \lambda_t\in \left(0,1\right] $. With this modification included, we derive the following lower bound to the end-to-end rate $R(L)$, again, while optimizing over $m$ (time-multiplexing block length) and $n$ (number of QR nodes).
	\begin{theorem}
		The rate-vs.-distance lower bound $R^{\text{LB}}(L)$ for a repeater chain with a lossy switching scheme (described by Eq.~\eqref{eqn:switch_modif}) has two regimes of operation described by the equation: 
		\begin{align}
			R^{\text{LB}}_{\text{lossy}}=\frac{(M\mu)^{\log_2\lambda_t+1}}{q\tau}\exp\left(-c_{\text{exp}}\,\alpha L - 2c_{\text{sub}}\sqrt{\alpha L}\right),
		\end{align} 
		with the constants $ c_{\text{exp}} $ and $ c_{\text{sub}} $ defined by: 
	\begin{subequations}
			\begin{align}
			c_{\text{exp}}&=-\log_2 \lambda_t, \, \\{\text{and}} \nonumber\\
			c_{\text{sub}}&=\left[\log\left(\frac{(M\mu)^{1+\log_2\lambda_t}}{q(1-1/e)}\right)^{\log_2 \lambda_t}\!\!\!\!-\log(q(1-1/e))\right]^{1/2}. \label{eq:swloss_const}
		\end{align}
	\end{subequations}
		\label{theorem:switch_loss}
	\end{theorem}
	
	\begin{proof}
		See Appendix~\ref{app_subsec:mod_lower} for a detailed proof.
	\end{proof}
	In the regime of low loss, i.e.\ $ \lambda_t\geq10^{-0.2} (\approx 0.631)$, the subexponential nature of the bound dominates and we have a reliable design that beats $R_{\rm direct}(L)$ above a certain range $L$. In fact, it is evident from Eq.~\ref{eq:swloss_const}, that for extremely efficient switches with no loss i.e.\ $ \lambda_t=1 $, $ c_{\text{exp}}\rightarrow 0 $ and $ c_{\text{sub}} \rightarrow \left[\log(1/q(1-1/e))\right]^{1/2} $, which yields the original lower bound in Eq.~\eqref{eq:lowerbound}. In the regime of high switching loss ($\lambda_t <10^{-0.2}$), the exponential nature of the lower bound dominates and there may be certain transmission lengths and switching-loss regimes where the lower bound drops below $R_{\rm direct}(L)$. An exact rate-vs.-distance calculation would be needed to verify the viability of repeater operation in such an operational condition.
	
	\begin{figure}[h!]
		\centering{\includegraphics[width=\linewidth]{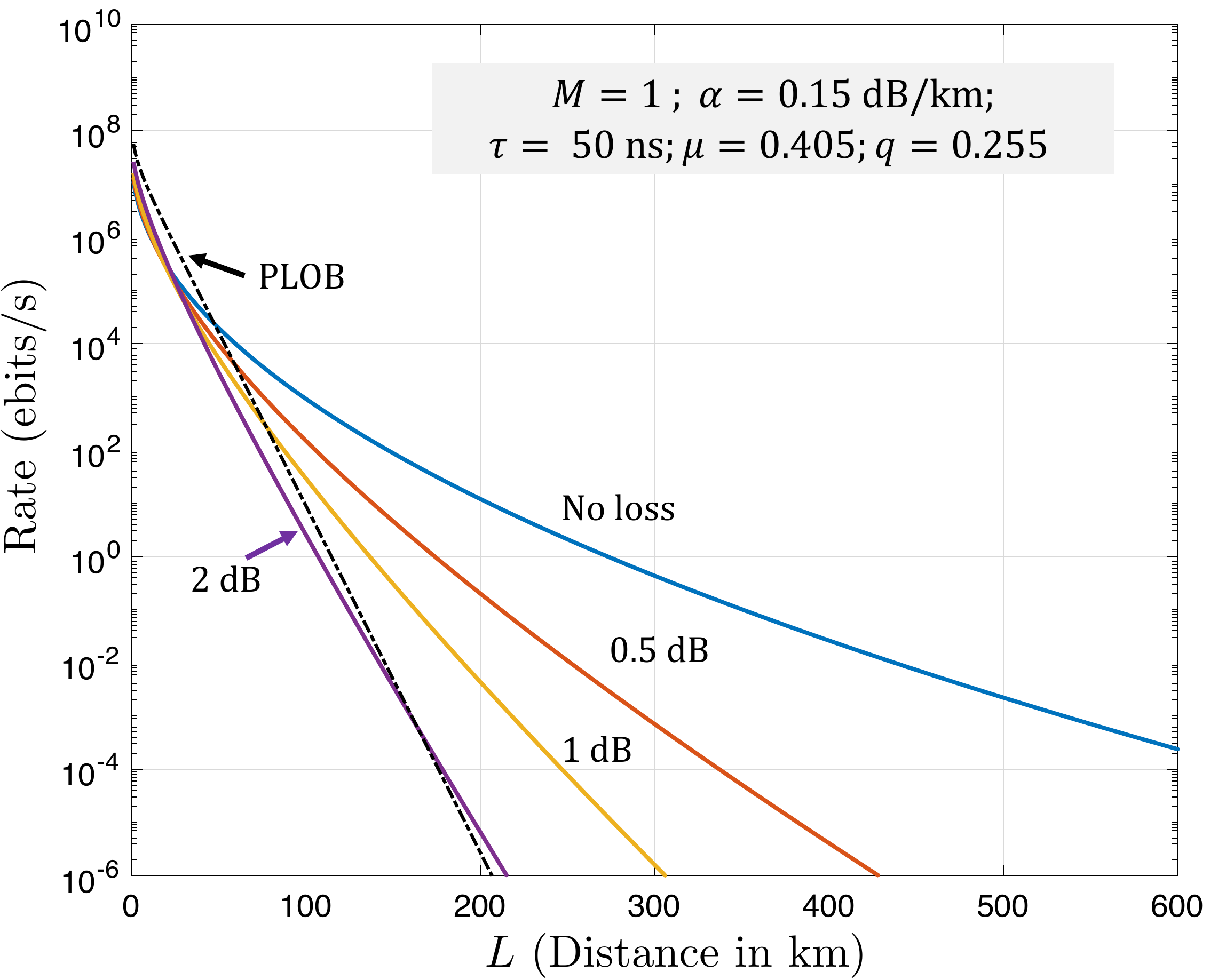}  }	
		\caption{Behavior of $R^{\rm LB}_{\text{lossy}}$ with increasing level of switching loss (marked on the plots). Only temporal multiplexing is employed, i.e., $M=1$. We assume $\alpha \equiv 0.15 \text{ dB/km}, \tau= 50 \text{ ns}, \mu=0.405, q=0.255$. As given by Eqn. \eqref{eq:swloss_const}, the subexponential rate-vs.-distance scaling (blue curve) is affected by the switching losses until the rate law becomes exponential and the design is unable to exceed the PLOB bound (in this case at $\lambda_t$ corresponding to 2 dB the advantage is lost). }
		\label{fig:M1}
	\end{figure}

	\begin{figure}[h!]
		\centering{\includegraphics[width=\linewidth]{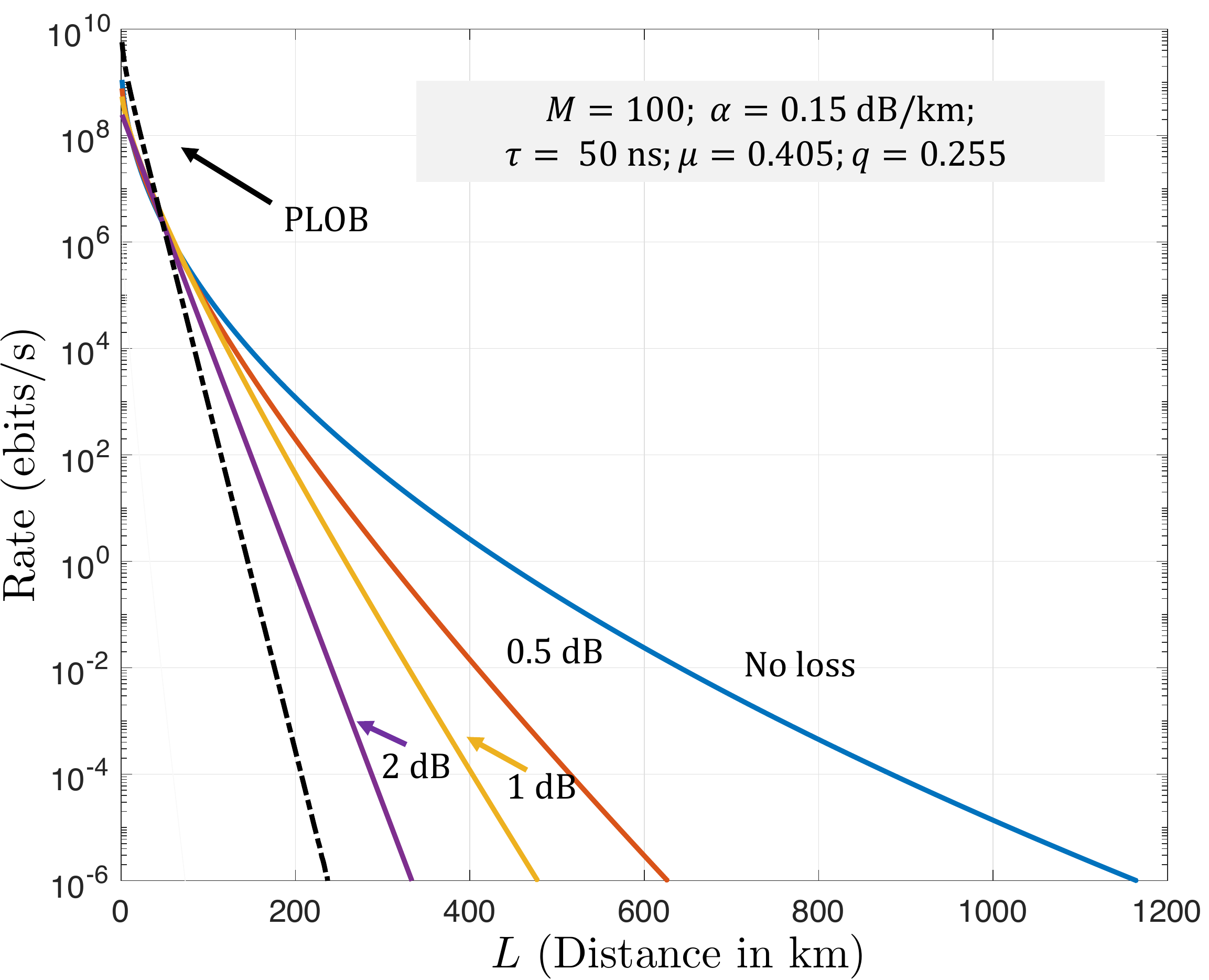}  }	
		\caption{Behavior of $ R^{\text{LB}}_{\text{lossy}} $ with increasing switching losses as marked. $ M=100,\, \alpha=0.15 \text{ dB/km}, \tau= 50 \text{ ns}, \mu=0.405, q=0.255 $. As given by Eq. \eqref{eq:swloss_const}, the subexponential rate-vs.-distance scaling (blue curve) is affected by the switching losses until the rate law becomes exponential.}
		\label{fig:M100}
	\end{figure}
We plot $ R^{\text{LB}}_{\text{lossy}} $ for $M=1$ in Fig.~\ref{fig:M1}, and for $M=100$ in Fig.~\ref{fig:M100}. In both cases we observe levels of switching loss (2 dB and 5 dB respectively) beyond which the rate scaling fails to surpass $R_{\rm direct}(L)$. Therefore given a set of device metrics, there is a maximum switching loss that is tolerable, beyond which entanglement distribution at rates surpassing $R_{\rm direct}(L)$ is unattainable. 

The derivation of $ R^{\text{LB}}_{\text{lossy}} $ also yields the following required optimal values for the number of repeaters ($ n $) and time multiplexing block length $ (m) $.
	
	\begin{theorem}
		Given the link distance between Alice and Bob, the optimal number of repeater stations $ {n}^*(L) $ and order of time multiplexing $ {m}^*(L) $ required to attain the lower bound, are given by the following formulas:
		\begin{align}
			&{n}^*(L):=\sqrt{\frac{\log_2 \lambda_t +1}{\log(M\mu)\log_2 \lambda_t-\log(q(1-1/e))}}\times \sqrt{\alpha L}  -1,\\
			&{m}^*(L):=\frac{\exp\left(\frac{\alpha L}{n^*(L)+1}\right)}{M \mu}.
		\end{align}
		\label{theorem:optimal_param}
	\end{theorem}
	
	\begin{proof}
		See Appendix~\ref{app_subsec:opti_vals} for a detailed proof.
	\end{proof}

	\begin{figure}[h]
		\centering
		\includegraphics[width=\linewidth]{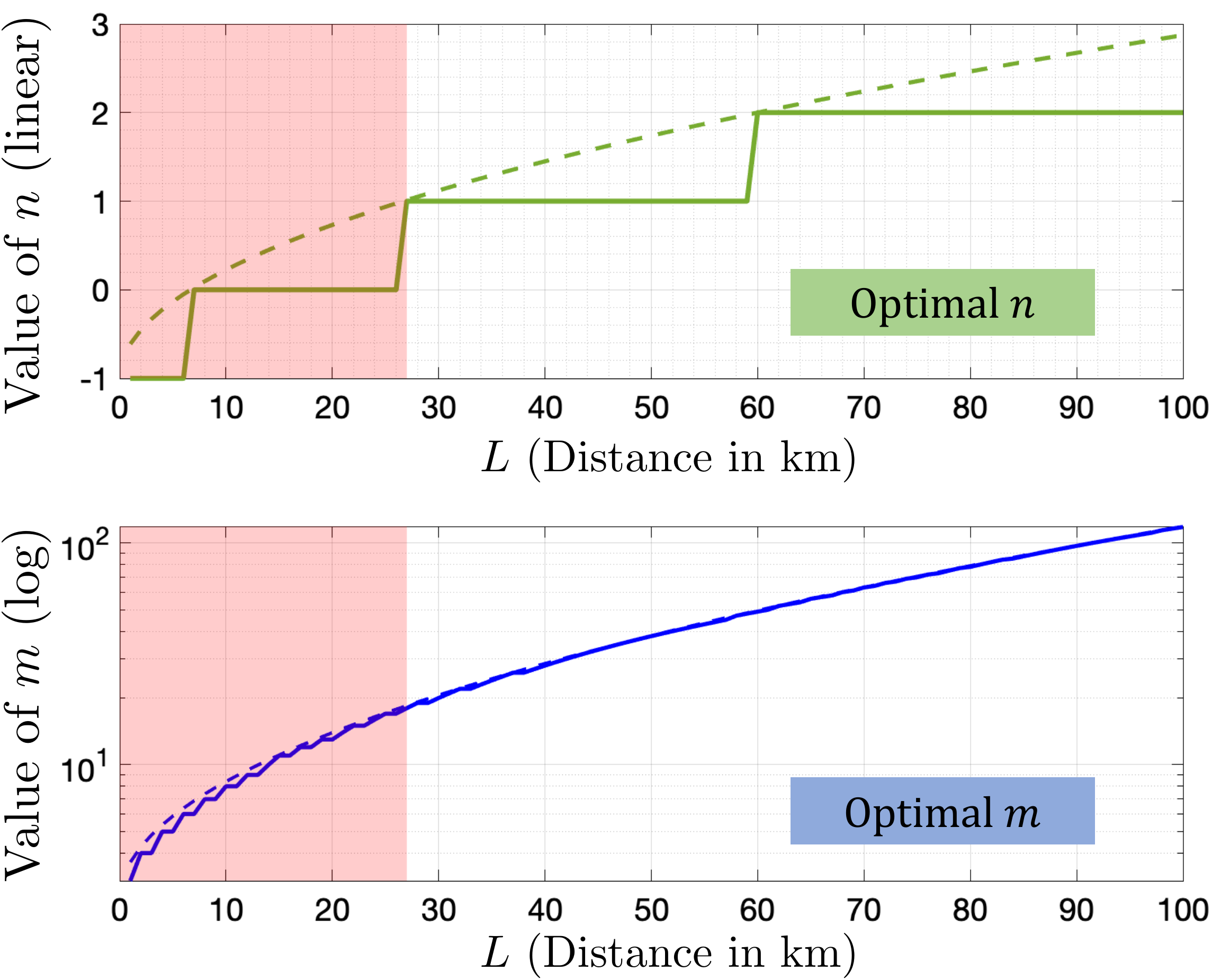}  
		\caption{Optimal values of $ m $ and $ n $ at a given distance with pure time multiplexing $ (M=1) $. Solid lines show integer values which are a floor of the analytic values. The red region represents lengths upto which a repeater protocol is not possible i.e. the number of repeaters is less than 1. }
		\label{fig:optim_param}
	\end{figure}

	\begin{figure}[hbp]
	\centering
	\includegraphics[width=0.95\linewidth]{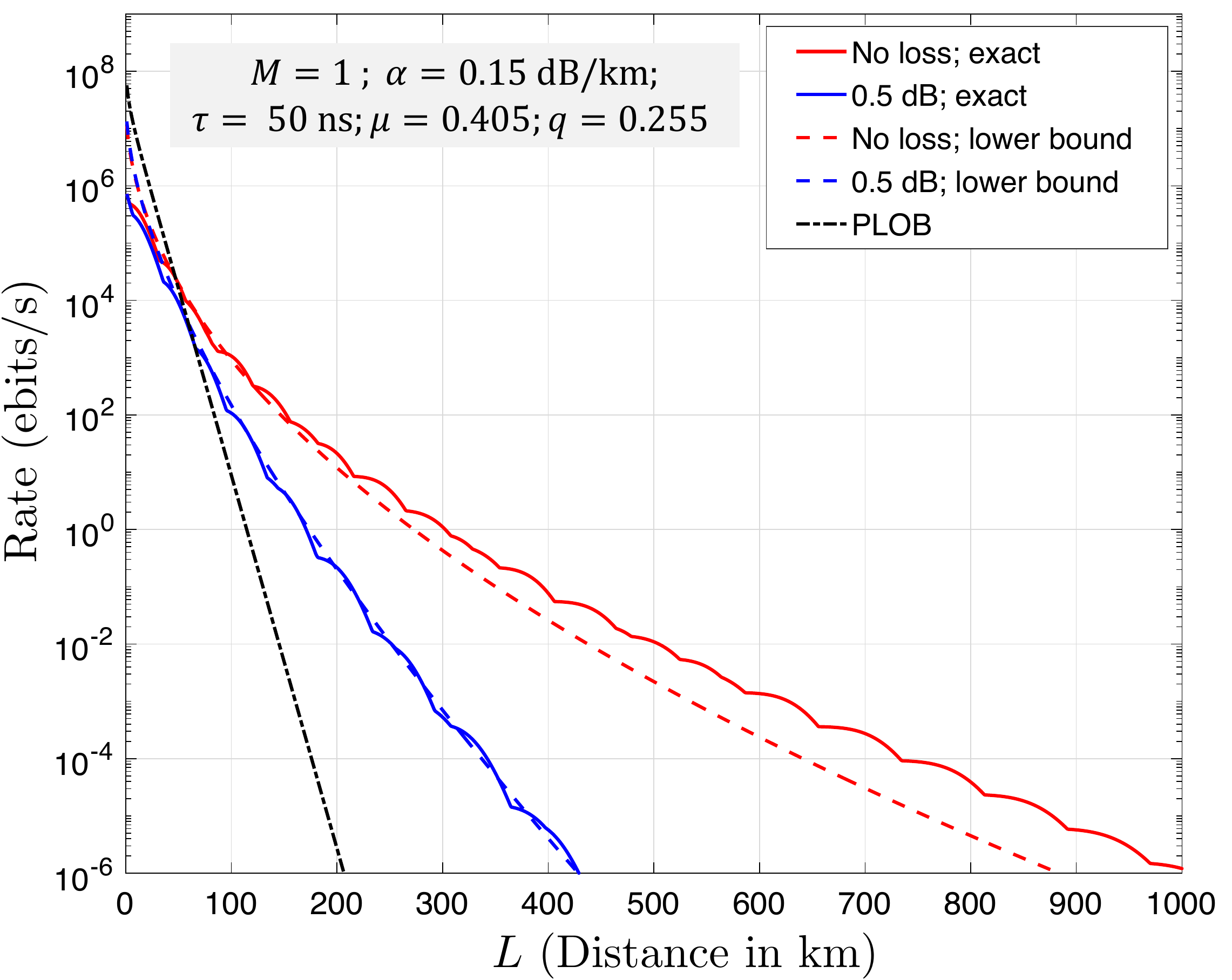}  
	\caption{Comparison of numerical rate envelope (solid lines) with the analytic lower bound (dashed lines). Note that till a certain distance the lower bound is over-optimized; as a repeater chain cannot be formed for the optimized values of $ n $ (repeater stations) and $ m $ (degree of time multiplexing).}
	\label{fig:comparison}
\end{figure}

In order to find the optimal values of $ n $ and $ m $, we pretended as if they were continuous parameters and took derivatives of the rate with respect to them. Yet, fractional quantities for $n$ and $m$ are physically meaningless. By considering the integer floor of continuous optima $n^*(L)$ and $m^*(L)$, we gain further insight into regions where optimal performance is not truly achievable, as observed in Fig.~\ref{fig:optim_param} for the purely time multiplexed repeater ($ M=1 $). The red-shaded region is where the number of repeaters is sub unity which has no physical meaning. This range of operation is therefore `forbidden'. From the rate plot in Fig.~\ref{fig:comparison}, the red region is under the repeater-less bound, which means that direct transmission would be superior to using our QR in this regime.

	\section{Accounting for Decoherence in Quantum Memories}
	\label{section:decoherence}
	
	Practically, the qubits stored in the QM register at the QR station will undergo a certain amount of loss proportional to the time that it must be held before the QM BSM is attempted. In general, we may understand it as a decoherence phenomenon modeled by some parameter, which we label $ \lambda_{\text{mem}} $ here. Similar to $\lambda_t$, we define $\lambda_{\text{mem}}$ as the probability that the qubit is preserved in the QM over a single time step $\tau$. In order to understand how decoherence affects the optimal rate, we consider the `worst case scenario' that qubits stored in QMs always undergo maximum decoherence proportional to $m\tau$ time slots, irrespective of the actual duration of time within the time-multiplexing block they actually lived in the QM. This leads to further modification of the QM entanglement swap success parameter as: 
	\begin{align}
		q\rightarrow q\times \lambda^{\log_2 m}_t \times \lambda_{\text{mem}}^m,
		\label{eqn:mem_modif}
	\end{align} 
	where $ \lambda_t, \lambda_{\text{mem}}\in \left(0,1\right] $.
	The additional modification changes the nature of the envelope as summarized in Theorem \ref{theorem:decoherence}, 
	\begin{theorem}
		The rate vs. distance lower bound $ R^{\text{LB}}(L) $ for a repeater chain with a loss prone switching scheme and quantum memory decoherence (described by Eq.~\eqref{eqn:mem_modif}) is given by:
		\begin{widetext}
			\begin{align}
				R^{\text{LB}}_{\text{loss,decoh.}}=\frac{(M\mu)^{\log_2\lambda_t+1}}{q\tau} \exp\left[-\alpha L \log_2\lambda_t\left(\frac{1}{v_0} -1\right) + 
				\frac{(-\alpha L)}{v_0}+v_0 \log\left(\frac{q(1/e-1)}{(M\mu)^{\log_2 \lambda_t}}\right)+\frac{e^{\alpha L/v_0}\log(\lambda_{\text{mem}})}{M\mu}\right],
			\end{align} 
		\end{widetext}
		where $ v_0 $ is the solution of the transcendental equation
		\begin{align}
			\log x \left[\frac{x^{-1/v}\log \lambda_{\text{mem}}}{M \mu}-\log_2 2\lambda_t\right] =v^2 \log \left(\frac{(M\mu)^{\log_2\lambda_t}}{q(1-1/e)}\right).
		\end{align}
		Note that the solution $ v_0 $ is distance $ (L) $ dependent.
		\label{theorem:decoherence}
	\end{theorem}

	\begin{proof}
		The proof proceeds similar to the proof of Theorem~\ref{theorem:switch_loss} (Appendix~\ref{appendix:switchloss}).
	\end{proof}
	
		\begin{figure}
		\centering
		\includegraphics[width=0.9\linewidth]{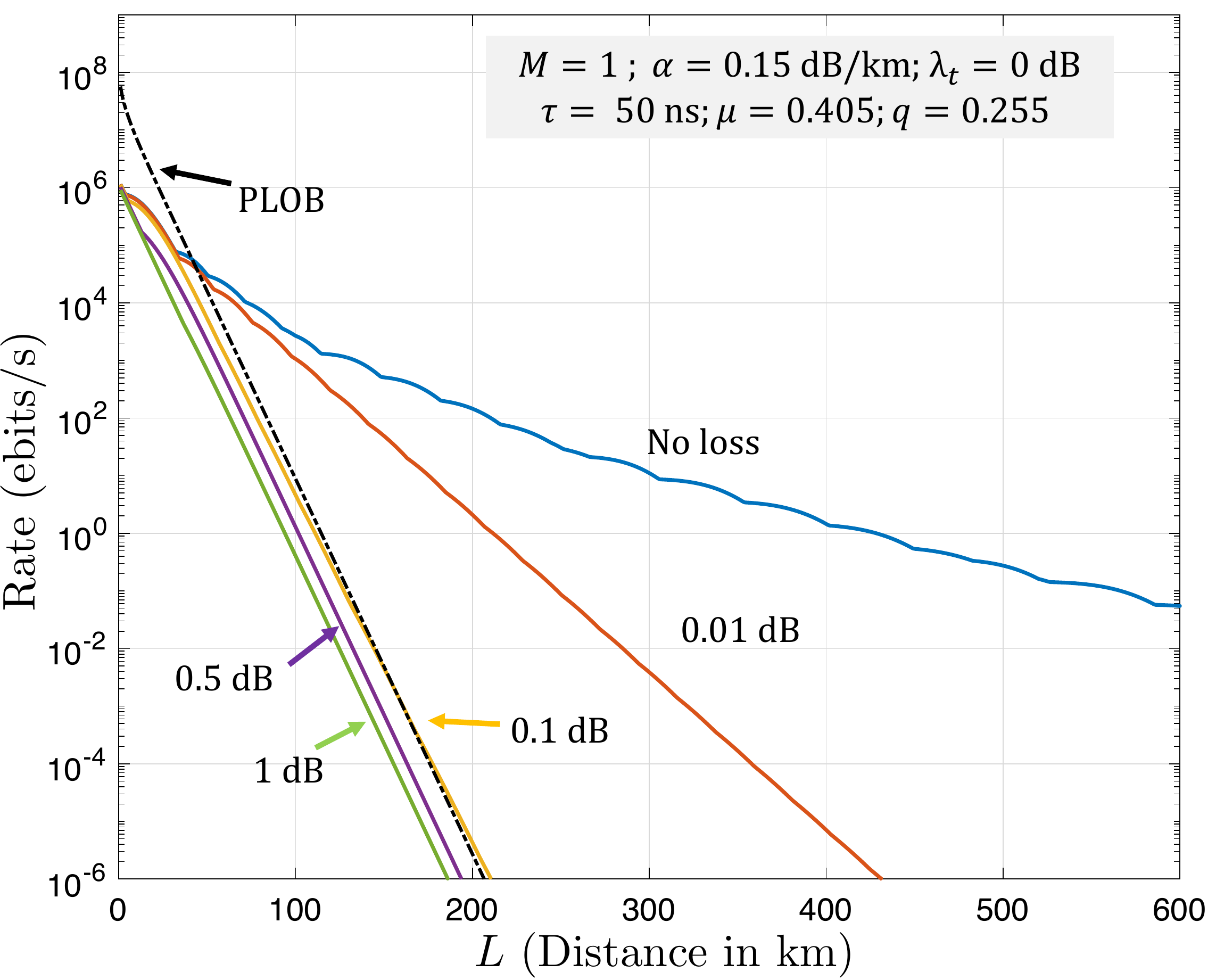}  
		\caption{Modification of rate-vs.-distance envelope lower bound at different values of $ \lambda_{\text{mem}} $ as specified; no switching loss considered.  $ M=1,\, \alpha=0.15 \text{ dB/km}, \tau= 50 \text{ ns}, \mu=0.405, q=0.255 $ }
		\label{fig:decoh1}
	\end{figure}
	As noted in Theorem~\ref{theorem:decoherence}, the distance dependent parameter $ v_0 $ prevents us from making direct conclusions about the scaling law of the rate-vs.-distance envelope. We choose to perform our analysis numerically and report numerical results to gain insights into the effect, as shown in Fig.~\ref{fig:decoh1}. We observe that the subexponential advantage is lost even for relatively small values of memory decoherence factor $ \lambda_{{mem}} $. This highlights that quantum memories with very high coherence times (much greater than the time required for one iteration of the protocol, $m\tau$) are required to maintain the subexponential advantage of time multiplexing. The availability of  highly durable QMs is therefore a necessity for any practical implementation of time multiplexed QRs.
	
	\section{Protocol Dependent Decoherence in Quantum Memories}
	\label{section:avg_decoh}
	As discussed in Sec.~\ref{section:bounds}, the actual effect of QM decoherence is event-dependent. The amount of decoherence will be governed by our strategy for mixing and matching the successful swap events spread over $ m\tau $ seconds. For example, the scheme that performs BSMs at the QRs at the end of $m$-th time step will experience more qubit decoherence than the scheme in which every QR performs BSM as soon as it has at least one link success on both sides. Based on the choice of the scheduling of QM-register BSMs, the time that a qubit is stored (and therefore the amount of decoherence) is now a random number. We model the general modification of the BSM success probability as follows,
	\begin{align}
		q\rightarrow q\times \lambda^{\log_2 m}_t \times \lambda_{\text{mem}}^{Y_i}\,,
	\end{align}
	where the wait time before BSM is performed at the $i{\text{th}}$ QR, is given by a random variable $Y_i$. The distribution for any $ Y_i $  depends on the number of time steps that red (i.e.\, stored) qubits have to wait before undergoing BSM.  Since link generation is probabilistic, we shall use two random variables $ X_{i,L} $ and $ X_{i,R} $, to denote the temporal location of a certain successful link generation to the left and right of the $ i{\text{th}}$ QR, respectively. The  functional dependence of $ Y_i $ in terms of  $ X_{i,L}  $ and $ X_{i,R} $ depends on the protocol rules that govern the QM switching and BSM scheduling.  For our present analysis, we consider two complementary protocols (exact descriptions and derivations of $ \langle Y_i\rangle $ for both protocols is given in Appendix \ref{appendix:avgloss}):
	\begin{enumerate}
		\item \textbf{Perform swap on the QMs corresponding to first success on each side:} We perform BSM on the QM immediately after both sides of a repeater station report their first successful links. For this protocol, $Y_i= |X_{i,L}- X_{i,R}| $.

		\item \textbf{Perform swap on QMs that have waited the least amount of time at the end of the length $ m $ time block:} In this case, we perform the BSM for the QMs only after all $ mM $ link generation attempts results are received. Subsequently, we choose to do the entanglement swap on the \textit{latest} successful link on either side of the QR. For this protocol, $ Y_i= X_{i,L}+X_{i,R}$.
		
	\end{enumerate}
	We note that, for both protocols $ X_{i,L/R} $ are independent and identically distributed (i.i.d.) random variables drawn from geometric distributions with the probability of success given by the probability of heralding at least a single successful LO Bell swap  i.e.\ $ P $ from Eq.~\eqref{eq:P_single}.
	
	\begin{figure}
		\centering
		\includegraphics[width=\linewidth]{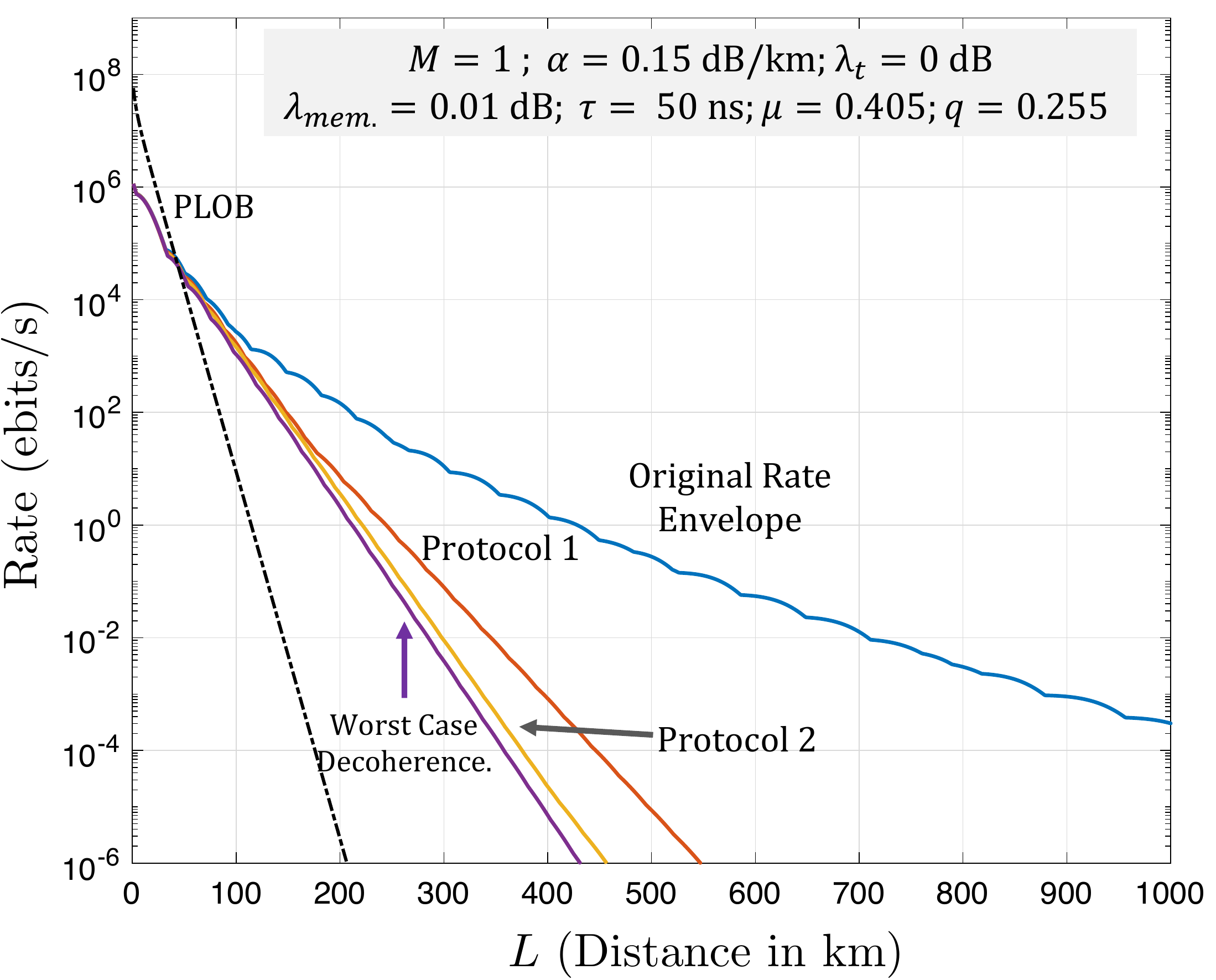}  
		\caption{Comparison of lower bounds to the achievable rates for the described average decoherence models with the subexponential lower bound and worst case decoherence model. }
		\label{fig:decoh2}
	\end{figure}
	
	As shown in Fig. \ref{fig:decoh2}, the lower bounds for the modified protocols surpass the rates achieved by considering maximum decoherence in QMs on an average. It is straightforward to see that protocol (1) beats protocol (2), because of the fact that only one red qubit has to wait for the QM BSM in the latter scenario. The optimal protocol would be able to surpass the performance of protocol (1). However this is not proven rigorously in this paper. All three models are shown to reliably surpass the PLOB bound for our choice of parameters, which supports their implementation for long distance entanglement generation.

	\section{Discussion and Conclusions}\label{sec:conclusions}
	Entanglement distribution at rates that surpass the direct transmission capacity is of paramount importance for the development of a quantum network. In the present study, we have analyzed the significance of time multiplexing for a linear-chain two-way quantum repeater that uses BSMs and switches, but no error correction, to achieve a fundamentally superior rate compared with a similar scheme that relies solely on multiplexing across parallel entanglement attempts, e.g., by multiplexing across spatial or frequency modes. We evaluated upper and lower bounds to the optimum rate-vs.-loss envelope attained by this architecture in its full generality, and showed that the achievable rate $ R(L)$ is proportional to $ R\propto A \exp(-s\sqrt{\alpha L}) $, where $ L $ is the end-to-end range in optical fiber, and $ A $ and $ s $ are constants that depend upon various device parameters. We showed that temporal multiplexing by itself is capable of achieving the aforesaid scaling law, with no spatial or spectral multiplexing. With only spatial or spectral multiplexing (parallel channels), quantum repeaters that use multiplexed BSMs can attain $ R\propto A \exp(-s\alpha L) $ with $s < 1$, whereas this can be attained with $s=1$ when no repeaters are employed (direct communication).
	
Practical considerations of a lossy switching network and decoherence in the quantum memory (QM) register at the repeater nodes are also included in our analysis. We describe regimes where these practical effects have a detrimental effect on the theoretically achievable rate. Including the effect of loss in switching networks incurs a transition from the aforesaid subexponential rate-vs.-distance scaling to an exponential rate-vs.-distance scaling law (with increasing switching losses). In fact, for a given choice of design parameters, with switching losses above a certain threshold, including temporal multiplexing may deteriorate the performance, e.g., of a spatially-multiplexed quantum repeater architecture. We observed that including even a small amount of QM decoherence affects the subexponential rate scaling advantage afforded by time multiplexing. However, given the actual amount of decoherence, some actual performance benefit over a non-time-multiplexed protocol may occur. 
	
A complete analytical study of two-way quantum repeaters along the lines of our work, but that accounts for the use of multiple successful connections between elementary links, and optimal entanglement purification and distillation at link and/or multi-hop levels, is left for future work. Additionally, a full system level analysis of the protocol with realistic Bell state generation devices, quantum memory decoherence models and stochastic models for infidelity caused by the staggered waiting times in the QM registers would be useful to evaluate the performance of our protocol in a practical deployment. We envision practical time-multiplexed repeaters with multiple success utilization to employ block distillation either at the user nodes or at the level of the QR stations, or both, to attain the desired Fidelity of the end-to-end entanglement delivered. Furthermore, we hope that the fundamental benefit offered by time multiplexing demonstrated in this paper can be leveraged for future quantum repeater and router designs for use in quantum networks with more complex topologies. 
	
	\section*{Acknowledgments}
	P.D. and S.G. would like to acknowledge funding support from L3Harris Technologies, under the Contract number A000483213, to develop the detailed analysis of time-multiplexed quantum repeaters. A.P. was supported by the National Science Foundation (NSF) Engineering Research Center for Quantum Networks (CQN), Grant No. 1941583. H.K. and S.G. developed the subexponential scaling law associated with time multiplexing, funded by the DARPA Quiness program Raytheon-BBN Subaward Contract No. SP0020412-PROJ0005188, under Northwestern University Prime Contract No. W31P4Q-13-1-0004, in 2016. S.G. and H.K. would also like to acknowledge useful discussions with Zachary Dutton, Christoph Simon, and Wolfgang Tittel.

	\appendix
	\section{Derivation of subexponential upper and lower bounds}
	\label{appendix:bound}
\begin{figure}[h!]
		\centering
		\includegraphics[width=\linewidth]{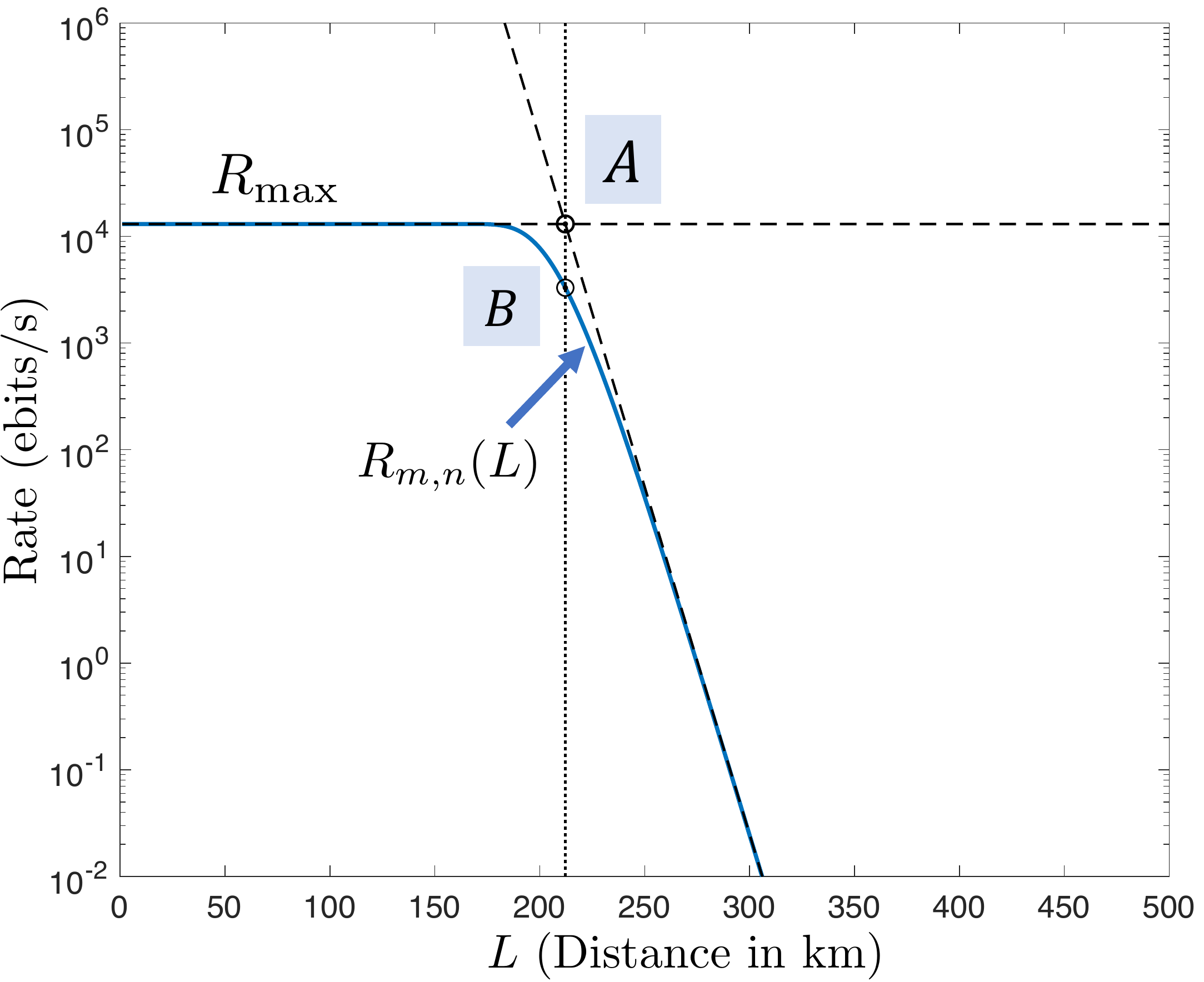}  
		\caption{The two-part upper bound to determine the family of points for deriving the bounds. The blue line is the actual rate plot for a fixed value of $ m $ and $n $. The dashed lines are the two-part envelope discussed. The marked points $ A $ and $ B $ are the points used to derive the upper and lower bounds respectively.}
		\label{fig:twopart}
	\end{figure}

	We begin by plotting the rate based on the exact rate equation in Eq.~\eqref{eq:rateeqn} for a fixed value of $ m $ and $ n $, in Fig.~\ref{fig:twopart}. Subsequently, we observe a two part upper envelope to the exact rate, which can be used to draw inferences about bounds to the rate-vs.-distance envelope. 
	The first segment of the two part envelope is given by the dashed line marked as $ R_{\text{max}} $, which can be expressed as,
	\begin{align}
		R_{m,n}(L)\leq \frac{q^n}{m \tau} :=R_{\text{max}},  
	\end{align} with equality for $ \mu e^{-\alpha L/(n+1)}=1 $.  
	
	For the sake of brevity, we make the substitution $ p=\mu e^{-\alpha L/(n+1)} $. The Taylor series expansion of the $ 1-(1-p)^{Mm}  $ component from Eq.~\eqref{eq:rateeqn} yields the following inequality,    
	\begin{align}
		1-(1-p)^{Mm} < 1-(1-Mmp)= Mmp.
	\end{align} 
	Therefore, this allows us to determine an exponential scaling law given by $y= \left[ {(Mmp)^{n+1}q^n}\right]/\left({m \tau}\right) $, which forms the second half of the two part envelope, marked as $ R_{\exp} $ in Fig.~\ref{fig:twopart}. The two upper bounds have a single parametrized point of intersection which is described by the family of points $ A(m,n) $, given as 
	\begin{align}
		A(m,n):= \left[(Mm\mu)^{-(n+1)},\; \frac{q^n}{m \tau}\right].
	\end{align}
	The locus of $ A(m,n) $ is determined by eliminating the two parameters, i.e.\ $ m $ and $ n $. We first eliminate $ m $ from the ordinate and abscissa of the family as follows
	\begin{align}
		m=\frac{x^{-1/(n+1)}}{M\mu}= \frac{q^n}{y\tau}.
		\label{eq:lowbnd_deriv}
	\end{align}
	Subsequently, the elimination of $ n $ by using $ n+1= \sqrt{\frac{\ln x}{\ln q}}$, yields the upper bound 
	\begin{align}
		R^{\text{UB}} (L):=\frac{M \mu}{q \tau} e^{-\left(2 \sqrt{\log \left(1 / q\right)}\right) \sqrt{\alpha L}},
		\label{eq:UB_deriv}
	\end{align}
	where, $ -\alpha L= \log x $. 
	
	A lower bound to the envelope  is obtained by tracking the locus of the point on the rate curve, which shares the same ordinate as the family $ A(m,n) $. We obtain this by dropping a vertical intercept from $ A(m,n) $ to the actual rate curve. Let us label this parametrized point  $ B(m,n) $, which is defined by,
	\begin{align}
		B(m,n):= \left[(Mm\mu)^{-(n+1)},\; \frac{q^n \left[1-(1-\frac{1}{Mm})^{Mm}\right]^{n+1}}{m \tau}\right].
	\end{align}
	We make a further simplification to this family, by using the observation that $ \forall\; m,M\in \mathbb{Z^+} $,
	\begin{align}
		\left(1-\frac{1}{Mm}\right)^{Mm}>1/e. 
	\end{align}
	This means that the following family (let us label it $ B'(m,n) $) is also a lower bound of the actual rate envelope, and is expressed as,
	\begin{align}
		B'(m,n):=\left[(Mm\mu)^{-(n+1)},\;\frac{q^n (1-1/e)^{n+1}}{m\tau}\right].
		\label{eqn:app_lowerbound}
	\end{align}
	The similarity of $ B'(m,n) $  to the family $ A(m,n) $ leads to the intuitive argument that the envelope for $ B'(m,n) $ has the same form as Eqn.~\eqref{eq:UB_deriv}, with the only modification being that $ q $ is replaced by $ q(1-1/e) $. The detailed derivation would follow a similar procedure as we have done for $ A(m,n) $ above. Hence, the lower bound can be expressed as,
	\begin{align}
		R^{\text{LB}}(L):=\frac{M \mu}{q \tau} e^{-\left(2 \sqrt{\log \left(1 / q\left(1-1/e\right)\right)}\right) \sqrt{\alpha L}}.
		\label{eq:LB_deriv}
	\end{align}

	\section{Incorporating the Effects of Switching Loss in Quantum Memory Entanglement Swap}
	\label{appendix:switchloss}
	
	\subsection{Modified lower bound derivation}
	\label{app_subsec:mod_lower}
	We showed  how to derive the family of points that serve as bounds to the rate-vs.-distance envelope, in Appendix~\ref{appendix:bound}. 
	We consider the model of switching loss discussed in Section~\ref{section:switching}. To include the effect of switching losses, the QM BSM probability of success $ q $ is modified as follows:
	\begin{align}
		q\rightarrow q \times \lambda^{\log_2 m}_t.
	\end{align}
	The parameter $\lambda_t$ quantifies the per-switch loss, which corresponds to $10\log_{10}(1/\lambda_t)$ dB of loss. This can be thought of as the probability with which a single switch succeeds in switching a photon (as opposed to losing it). 

	Approaching this envelope calculation with a similar technique as was used for the original lower bound (\ref{eq:lowbnd_deriv}), the present lower bound abscissa can be expressed as
	\begin{align}
		y&=\frac{q^n (1-1/e)^{n+1}\times  \lambda^{n\log_2 \left({x^{-1/(n+1)}}/{(M\mu)}\right)}_t}{{\tau \times x^{-1/(n+1)}}/{(M\mu)} },
	\end{align} where $ x\equiv L;\, y\equiv R $. To keep track of the exponents efficiently, we recast the equation as
	\begin{widetext}
		\begin{align}
			\log y -\frac{1}{n+1}\log x=\frac{-n}{n+1}\log x \log_2 \lambda_t-n \log(M\mu)\log_2 \lambda_t+(n+1)\log (1-1/e)+n\log q +\log\left( \frac{M\mu}{\tau}\right).
			\label{eqn:fam2}
		\end{align}
	\end{widetext}
	This is a general parametrized family of curves in the two variables $ x $ and $ y  $, with $ n $ as the parameter we seek to eliminate. Let us denote this as an implicit function $ f(x,y,n) $. To determine the envelope of this family, the equations $ f(x,y,n)=0 $  and $ {\partial f(x,y,n)}/{\partial n} =0 $ must be solved simultaneously to eliminate the parameter $ n $~\cite{Bruce1992-si}. Partial differentiation of Eq.~\eqref{eqn:fam2} with respect to $ n $ yields,
	\begin{widetext}
		\begin{align}
			&\frac{1}{(n+1)^2}\log x \left(\log_2 \lambda_t+1\right)=\log(q(1-1/e))-\log(M\mu)\log_2 \lambda_t\\
			\Rightarrow &n+1=\left( \frac{\log x \left(\log_2 \lambda_t+1\right)}{\log(q(1-1/e))-\log(M\mu)\log_2 \lambda_t}\right)^{1/2}.
			\label{eqn:fam21}
		\end{align}
	\end{widetext}
	To simplify the derivation and identify terms that are relevant to the envelope, we make the substitutions $ -\log x = u $ and $ \log y=v $. Therefore Eq.~\eqref{eqn:fam21} can be rewritten as $ n+1= c_0 \cdot u^{1/2} $. Here $ c_0 $ is defined as,
	\begin{align}
		c^2_0= \frac{\log_2 \lambda_t +1}{\log(M\mu)\log_2 \lambda_t-\log(q(1-1/e))} .
	\end{align}
	Making the suitable substitutions in Eq.~\eqref{eqn:fam2} yields
	\begin{widetext}
		\begin{align}
			v+\frac{u}{c_0 u^{1/2}}=&\left(1-\frac{1}{c_0} u^{-1/2}\right) u \log_2 \lambda_t -c_0 u^{1/2} \log (M\mu)\log_2 \lambda_t +c_0 u^{1/2}\log (1-1/e)\nonumber\\ &+ c_0 u^{1/2}\log q+\log(\frac{M\mu}{q\tau})+ \log (M\mu)\log_2 \lambda_t\\
			\Rightarrow v=& u \log_2 \lambda_t- 2u^{1/2}[\log_2 \lambda_t\left\{\log(M\mu)\log_2 \lambda_t+\log(M\mu)-\log(q(1-1/e))\right\}\nonumber\\
			&-\log(q(1-1/e))]^{1/2}	+\log (M\mu)\log_2 \lambda_t+\log\left(\frac{M\mu}{q\tau}\right)\label{final_loss}
		\end{align}
	\end{widetext}
	As a point of reconciliation, in the case where we have no switching loss i.e.\ $ \lambda_t=1 $, we have $ c_0=(-\log (q(1-1/e)) )^{-1/2}$ and Eq.~\eqref{final_loss} becomes the original lower bound in Eq.~\eqref{eq:LB_deriv}.
	Thus, in the most general form, the overall lower bound may be written as 
	\begin{widetext}
		\begin{align}
			R^{\text{LB}}_{\text{lossy}}=\frac{(M\mu)^{\log_2\lambda_t+1}}{q\tau}\exp\left(-c_{\text{exp}} \alpha L - 2c_{\text{sub}}\sqrt{\alpha L}\right),
		\end{align}  
		\begin{subequations}
			\begin{align}
				\text{ where, } \nonumber\\
				 c_{\text{exp}}&=-\log_2 \lambda_t,\\
				c_{\text{sub}}&=\left[\log\left(\frac{(M\mu)^{1+\log_2\lambda_t}}{q(1-1/e)}\right)^{\log_2 \lambda_t}-\log(q(1-1/e))\right]^{1/2} .
			\end{align}
		\end{subequations}
	\end{widetext}
	 The derivation for Theorem~\ref{theorem:decoherence} proceeds in a similar fashion to the above. 
	 	
	\subsection{Optimal parameter value extraction}
	\label{app_subsec:opti_vals}
	From the above parameter elimination process, we can determine analytic expressions for the optimal value of $ n $ and $ m $, for a fixed choice of network parameters (i.e.\ link length, efficiencies etc.). From Eqs.~\eqref{eq:lowbnd_deriv} and~\eqref{eqn:fam21}, the optimal values as extracted are given by,
	\begin{align}
		&\bar{n}(L):=\sqrt{\frac{\log_2 \lambda_t +1}{\log(M\mu)\log_2 \lambda_t-\log(q(1-1/e))}}\times \sqrt{\alpha L}  -1,\\
		&\bar{m}(L):=\frac{\exp\left(\frac{\alpha L}{\bar{n}(L)+1}\right)}{M \mu}.
	\end{align}

	\section{Model for Average Decoherence in Quantum Memories}
	\label{appendix:avgloss}
	
	As per the notation developed in Sec.~\ref{section:avg_decoh}, we use $ Y_i $ to denote the random variable (for the $ i $-th repeater node) that governs the amount of loss due to the linear memory decoherence model of the QM. Specific protocol choices for the QM entanglement swap will govern the actual effect of such a loss term on the overall rate scaling. In general, the swap is only performed after there is \textit{at least} one successful link established on \textit{both} sides of the repeater station. Since link generation itself is probabilistic, we shall use two random variables $ X_{i,L} $ and $ X_{i,R} $, to denote the temporal location of a certain successful link generation to the left and right of the repeater node, respectively. The criteria for which successful link is marked depends on the protocol. 
	
	Hence, in general $ Y_i $ is a function of the success markers i.e.\ $ X_{i,L/R} $. We restate the
	protocols here

	 \textit{(1) Perform swap on the QMs corresponding to first success on each side:} We perform BSM on the QM immediately after both sides of a repeater station report their first successful links. For this case $Y_i= |X_{i,L}- X_{i,R}| $ where $ X_{i,L/R} $  indicate the number of time steps the first entangled QMs on the left and right of $i{\text{-th}}$ QR  have waited for, until the BSM is performed.

		\textit{(2) Perform swap on QMs that have waited the least amount of time at the end of the length $ m $ time block:} In this case, we perform the BSM for the QMs only after all $ mM $ LO BSM attempts (on either side of the QR) results are received. Subsequently we choose to do the entanglement swap on the latest successful link on either side of the QR. Here, we use the random variables $ X_{i,L} $ and $ X_{i,R}  $ to display the wait times for the latest entangled qubit, i.e.\ ,the \textit{last} link that was successfully generated on the left and right of the $i{\text{-th}}$ QR, respectively.
	
	Our rate equations account for the memory decoherence in the form of a general exponential with a sub-unity base (i.e. $\lambda_{\text{mem}}\in (0,1] $). Since $ f(x)=p^x $ is a convex function for $ p\in(0,1) $, we can use Jensen's inequality for a random variable $ X $, 
	\begin{align}
		\lambda_{\text{mem}}^{\langle X \rangle }\leq\langle \lambda_{\text{mem}}^{X}\rangle.
	\end{align}
	Hence, we can reliably use the mean value of the exponents to lower bound the mean of the general exponential function. This lower bound is sufficient to study the rate-vs.distance scaling for our modified protocols and make inferences about their corresponding performances.
	
	For protocol (1), $Y_i= |X_{i,L}- X_{i,R}| $ where $ X_{i,L/R} $ are drawn from geometric distributions with probability of success given by $ p $ i.e. $\Pr (X_{i,L/R}=k)=(1-p)^{k-1} p$. Here, the exponent for $ \lambda_{{mem}} $ in the final rate equation, is given by $ S= \sum_{i=1}^{n} Y_i $ with $ \langle S\rangle = n\Delta_1 $, where $ \Delta_1 $ is the mean value of the order statistics of the difference of two i.i.d. negative binomial distributions \cite{consul1989}. It can expressed in terms of the probability of success $ p $ as 
	\begin{align}
		\Delta_1= 2p^2 \sum_{j=1}^{\infty}(1-p)^{2j}\sum_{s=1}^{j} \frac{s}{(1-p)^s}=\frac{2(1-p)}{(2-p)p}.
	\end{align} 
	
	For protocol 2, to calculate the $ \Pr(X_{i,L/R}=k)$ i.e.\ probability that the last success has to wait for $ k $ time steps before the swap, we may adopt the following argument
	\begin{align}
		\begin{split}
			&\Pr(X_{i,L/R}=k)\\&=\Pr(\text{Atleast one success at the $ (m-k-1) $-th step})\\& \times \Pr(\text{No success for last $ k $ steps})\\& =(1-(1-p)^M) (1-p)^{kM}.
		\end{split}
	\end{align}
	It is easy to show that the expectation value $ \langle X_{i,L/R} \rangle $ is given by
	\begin{align}
		\langle X_{i,L/R} \rangle&=\frac{\sum_{k=0}^{m-1} k \times P(X_{i,L/R}=k)}{\sum_{k=0}^{m-1} P(X_{i,L/R}=k)}\\
		&=\frac{(1-p)^M}{1-(1-p)^M}-\frac{m (1-p)^{m M}}{1-(1-p)^{m M}}.
	\end{align}
	
	\nocite{*}
	\bibliography{bibliography}
\end{document}